\documentclass[aps, twocolumn, showpacs, table, longbibliography]{revtex4-1}
\usepackage[T1]{fontenc}
\usepackage[utf8]{inputenc}
\usepackage{graphicx}
\usepackage{pstricks}
\usepackage{array}
\usepackage{amsmath}
\usepackage{pifont}
\usepackage{dsfont}
\usepackage{multirow}
\usepackage{amssymb}
\usepackage{wasysym}
\usepackage[titletoc,toc,title]{appendix}
\usepackage{booktabs}
\usepackage{colortbl}
\usepackage{xcolor}
\usepackage{mathtools}
\usepackage{gensymb}
\usepackage{breqn}
\usepackage{sidecap}
\usepackage{footmisc}

\def\be{\begin{equation}}
\def\ee{\end{equation}}
\newcommand{\eq}[1]{Eq.~(\ref{#1})}

\def\GB{\mathrm{_{GB}}}

\newcommand{\cnst}{Center for Nanoscale Science and Technology, National
Institute of Standards and Technology, Gaithersburg, MD 20899, USA}
\newcommand{\umd}{Maryland NanoCenter, University of Maryland, College
Park, MD 20742, USA}

\begin{document}

\title{Charged grain boundaries and carrier recombination in polycrystalline
    thin film solar cells}

\begin{abstract}
We present analytical relations for the dark recombination current of a $pn^+$
junction with positively charged columnar grain boundaries in the high defect
density regime. We consider two defect state configurations relevant for
positively charged grain boundaries: a single donor state and a continuum of
both acceptors and donors. Compared to a continuum of acceptor+donor states, or
to the previously studied single acceptor+donor state, the grain boundary
recombination of a single donor state is suppressed by orders of magnitude. We
show numerically that superposition holds near the open-circuit voltage $V_{\rm
oc}$, so that our dark $J(V)$ relations determine $V_{\rm oc}$ for a given short
circuit current $J_{\rm sc}$. We finally explicitly show how $V_{\rm oc}$
depends on the these two grain boundary defect state configurations.
\end{abstract}

\author{Benoit Gaury}
\affiliation{\cnst}
\affiliation{\umd}
\author{Paul M. Haney}
\affiliation{\cnst}

\date{\today}

\maketitle

\section{Introduction}

{Polycrystalline photovoltaics have seen substantial increases in power
conversion efficiency in recent years, exceeding the 20~\%
mark~\cite{Green2017}.  Many of these advancements are due to improvements in
light management, and have resulted in a short-circuit current density $J_{\rm
sc}$ at 95~\% of its theoretical maximum in CdTe~\cite{geisthardt2015status}.
However, there remains substantial room for improvement in the open-circuit
voltage $V_{\rm oc}$; the current record value of 850~mV in CdTe is only 76~\%
of its theoretical maximum~\cite{geisthardt2015status}.  Despite the impressive
progress in thin film photovoltaics, fundamental questions regarding the role of
grain boundaries persist.  For example, the relatively low efficiency of single
crystal CdTe has led some to suggest that grain boundaries are beneficial to
photovoltaic performance~\cite{Jiang2004, visoly2006}, while numerical simulations typically
indicate that this is not the
case~\cite{gloeckler2005grain,taretto2008numerical,rau2009grain,
troni2013simulation}.  There are multiple reasons for the uncertainty regarding
grain boundaries: experimentally, grain boundaries are difficult to
independently control and measure, and theoretically there is not a simple
analytical model which fully captures the physics of grain boundaries' impact on
photovoltaic performance.

Two recent reports~\cite{Zhao2016, Burst2016} on single cyrstal CdTe solar cells
showed open-circuit voltages above 1~V, indicating that grain boundaries are a
primary source of $V_{\rm oc}$ losses due to recombination.  This finding has
renewed the impetus to understand and mitigate the impact of grain boundaries on
$V_{\rm oc}$.  Most previous theoretical works in this direction have consisted
of numerical
simulations~\cite{gloeckler2005grain,taretto2008numerical, rau2009grain,
troni2013simulation, Kanevce2017}. Alternately, analytical models offer a
concise, quantitative description of system behavior while providing further
insight.

In light of the need for improved understanding of grain boundaries' impact on
$V_{\rm oc}$, we recently developed an analytical description of dark grain
boundary recombination current, with the primary result of a simple relation
between grain boundary properties and $V_{\rm oc}$~\cite{Gaury2016GB}.  Some
details of the grain boundary model in this previous work are rather particular:
we assumed that the grain boundary defect states consist of a donor and an
acceptor at the same energy (so-called negative U-center).  This assumption has
been used in previous studies~\cite{edmiston1996improved, taretto2008numerical}
and is a simple way for the grain boundary to exhibit Fermi level pinning.
However, the performance of polycrystalline photovoltaics
like CdTe and Cu(In,Ga)Se$_2$ depends critically on the grain boundary defect
chemistry~\cite{rudmann2004, kumar2014physics}.  An adequately general model
should therefore accommodate varied defect spectra.

\bigskip 
In this work, we generalize our previous analysis to consider other grain
boundary defect configurations. We provide closed form expressions and physical
descriptions relating grain boundary properties to $V_{\rm oc}$. 
Because a majority of
experimental works provide evidence for positively charged grain boundaries in
CdTe~\cite{yoon2013local,moutinho2010investigation,tuteja2016direct},
 we restrict our attention to positively charged defects: a
 single donor defect and a continuum of donor and acceptor defects.  We found
 that generalizing the grain boundary defect configurations
necessitated a clearer and more general formulation of the model assumptions
given in Ref.~\onlinecite{Gaury2016GB}.
The scope of this analysis is limited to grains with interior
electrostatically similar to the unperturbed (i.e. grain boundary free) $pn$
junction, and materials with large intragrain hole mobility (on the order of
$50~\rm cm^2/(V\cdot s)$ which is consistent with single crystal
CdTe~\cite{Sellin2005, Duenow2014}). 
To provide specific context, we present much of our analysis in terms of
material parameters related to CdTe solar cells.  However, our analysis is not
material-specific, and applies to any material which conforms to the assumptions
we make.

There are qualitative similarities between the behavior of the single
donor+acceptor considered in Ref.~\onlinecite{Gaury2016GB} and the single donor
and continuum cases studied here.  In all cases, there are three regimes of
qualitatively distinct behavior, which can roughly be classified according to
the grain boundary core type: $n$-type, $p$-type, or neither (the latter case
applies at high applied voltages, where both electron and hole densities vary
with voltage).  The full explicit form of the grain boundary dark current is
shown in Table~\ref{form} for all defect configurations.  An important new
finding in this work is that compared to a continuum of acceptor+donor states,
or to the previously studied single acceptor+donor state, the grain boundary
recombination of a single donor state is suppressed by orders of magnitude.  

This article is structured as follows. We begin with a description of the
physical model for the grain boundary, encompassing both the single donor case
and the continuum of gap states in Sec.~\ref{model}. We summarize the
equilibrium properties of the grain boundary and our assumptions for analyzing
the out-of-equilibrium problem in Sec.~\ref{assumptions}. We present the charge
transport and associated grain boundary dark current for the single donor state
in Sec.~\ref{donor}, and for the continuum of gap states in
Sec.~\ref{continuum}. We end the article with Sec.~\ref{secVoc} where we discuss
the implications of our analysis on the open-circuit voltage of an illuminated
$pn$ junction. Finally, we examine the differences between the various
configurations of the gap states considered in the paper.

\begin{table}
\def\arraystretch{1.3}
\resizebox{0.49\textwidth}{!}{
\begin{tabular*}{0.6\textwidth}{lcccc}
  \toprule
\multicolumn{5}{c}{Grain boundary dark current: $J\GB(V)=\lambda \frac{S}{d} N e^{-E_a/k_BT} e^{qV/(nk_BT)}$}\\
\midrule
Defect(s) & Param. & $n$-type & $p$-type & high-recombination\\
  \midrule
  \multirow{5}{*}{single D}

                            &n & $1$ & $1$ & $2$\\
                            &$E_a$&  $E\GB$ & $E_g-E\GB$ & $E_g/2$\\
                            &N& $N_V$ & $N_C$ & $\sqrt{N_CN_V}$ \\
                            &S& $(1-f_0)S_p$ & $S_n$ & $\sqrt{(1-f_0)S_nS_p}$\\
                            &$\lambda$ & $L\GB$ & \begin{tabular}{cc}
                                $L\GB~\mathrm{for}~L_n \gg L\GB$ \\
                                $L_n~\mathrm{for}~L_n \ll L\GB$ \end{tabular}&
                                \begin{tabular}{cc}
                                $L\GB~\mathrm{for}~L_n' \gg L\GB$ \\
                                $L_n'~\mathrm{for}~L_n' \ll L\GB$ \end{tabular}
  \\
  \midrule
  \multirow{5}{*}{single A+D}
                              & n  & $1$ & $1$ & $2$\\
                              & $E_a$  & $E\GB$ & $E_g-E\GB$ & $E_g/2$\\
                              & N&  $N_V$ & $N_C$ & $\sqrt{N_CN_V}$ \\
                                &S&  $S_p/2$ & $S_n/2$ & $\sqrt{S_nS_p}/2$\\
                              & $\lambda$  & $L\GB$ & \begin{tabular}{cc}
                                $L\GB~\mathrm{for}~L_n \gg L\GB$ \\
                                $x_0~\mathrm{for}~L_n \ll L\GB$ \end{tabular}&
                                \begin{tabular}{cc}
                                $L\GB~\mathrm{for}~L_n' \gg L\GB$ \\
                                $L_n'~\mathrm{for}~L_n' \ll L\GB$ \end{tabular}
  \\
  \midrule
  \multirow{5}{*}{\parbox{1.5cm}{Continuum A+D}}
                              &n  & $1$ & $1$ & $2$\\
                              &$E_a$  & $E\GB$ & $E_g-E\GB$ & $E_g/2$\\
                              & N& $N_V$ & $N_C$ & $\sqrt{N_CN_V}$ \\
                               &S & $\mathcal{S}_p$ & $\mathcal{S}_n$ &
                               $\mathcal{S}/\sqrt{\gamma}$\\
                              &$\lambda$  & $L\GB$ & \begin{tabular}{cc}
                                $L\GB~\mathrm{for}~L_n \gg L\GB$ \\
                                $x_0~\mathrm{for}~L_n \ll L\GB$ \end{tabular}&
                                \begin{tabular}{cc}
                                $L\GB~\mathrm{for}~L_n' \gg L\GB$ \\
                                $L_n'~\mathrm{for}~L_n' \ll L\GB$ \end{tabular}
  \\

  \bottomrule\\
\end{tabular*}
}
\caption{\label{form} Summary of analytical results for the grain boundary recombination
current for various  defect density of states (single donor, single donor and
acceptor, continuum of donors and acceptors). The general form of the grain boundary
dark current is $J\GB(V)=\lambda S/d N e^{-E_a/k_BT} e^{qV/(nk_BT)}$ where
$S$ is a surface recombination velocity, $\lambda$ is a length characteristic of
the recombination region, $N$ is an effective density of states, $E_a$ is an
activation energy, $n$ is the ideality factor, $d$ is the grain size and $V$ is
the applied voltage.  Each column corresponds to the regime in which the grain
boundary is depending on voltage. $L\GB$ is the length of the grain boundary,
$L_n$ and $L_n'$ are effective electron diffusion lengths, $x_0$ is given by
\eq{x0}. $f_0$ is the thermal equilibrium occupancy of the single donor state
($f_0\approx 1$). $\gamma$ is given by \eq{gamma}. }
\end{table}

\section{Physical model of the grain boundary}
\label{model}
Our model system is depicted in Fig.~\ref{system}(a): a $pn^+$ junction of width
$d$ and length $L$ with a grain boundary perpendicular to the junction. We use
periodic boundary conditions in the $y$-direction so that the system describes a
closed grain of width $d$ ($d=3~\mu\rm m$ in our numerical calculations). 
We assume selective contacts so that the majority (minority) carrier surface
recombination velocity is infinite (zero), which implies that the electron (hole)
current vanishes at $x=L$ ($x=0$).
We define $x_0$ as the
position where electron and hole concentrations are equal in the grain interior.
As stated in the introduction we focus on
positively charged grain boundaries, which require screening by nearby negative
charges (free electrons or ionized acceptor dopants) to
conserve the device electroneutrality.  The consequences of the screening on the
electrostatics surrounding the grain boundary
depend on the statistics of the gap levels, the defect density of
states, and on the grain interior type ($n$-type or $p$-type region). For
example, a $p$-type material will develop an electric field around the grain
boundary to repel free holes from the grain boundary core, creating a depleted
region around it. Because of the absence of holes, this region has a negative
charge compensating the positive charge of the defect, as shown in
Fig.~\ref{system}(b). Electroneutrality determines the spatial extent of this
depleted region: the net charge (grain boundary charge plus depleted
charge) of the ensemble grain boundary/depletion region is zero.

The consequence of the aforementioned electric field on the local energy
landscape can be seen on a band structure plot across the grain boundary, as
shown in Fig.~\ref{system}(c). The electric field leads to the bending of
the conduction and valence bands, leading to a
built-in potential $V\GB$ around the grain boundary.
Depending on materials parameters and the extent of band bending, the grain
boundary core may undergo type inversion with respect to the grain interior.
Each grain boundary type has different carrier transport properties under
nonequilibrium conditions. In this work we consider both inverted and
non-inverted grain boundaries.

\begin{figure}
 \includegraphics[width=0.49\textwidth]{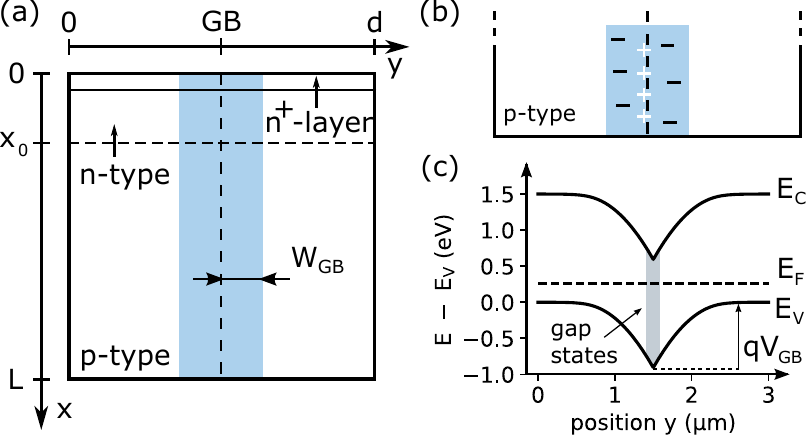}
 \caption{\label{system} (a) 2D model system of a $pn^+$ junction
 containing a columnar grain boundary. The depletion region of the grain boundary
  is indicated in blue (width $2W\GB$). $x_0$ is the point in the grain
  interior where electron and hole densities are equal. (b) Static charges at
  and around the grain boundary core ($+$ and $-$ signs respectively) in the
  $p$-type part of the system. (c) Schematic of a band structure corresponding
  to the situation described in (b).}
\end{figure}

\bigskip
The grain boundary is modeled as a two-dimensional plane with various
concentrations of donor and acceptor gap states. The grain boundary defect charge
density reads
\be
Q\GB = q \int_0^{E_g}\mathrm{d}E\ \rho_D(E)\left[1-f_D(E)\right] -
\rho_A(E) f_A(E),
\label{QGB}
\ee
where $\rho_D$ ($\rho_A$) is a two-dimensional density of donor (acceptor)
states per energy unit. The occupancies of each donor/acceptor state (indicated
by the index $k \in (D,A)$) is~\cite{PhysRev}
\be
    f_k(E) = \frac{S^k_n n\GB + S^k_p \bar
    p\GB(E)}{S^k_n(n\GB+\bar n\GB(E)) + S^k_p(p\GB+\bar p\GB(E))},
    \label{eq:fGBe}
\ee
where $n\GB$ ($p\GB$) is the electron (hole) carrier density at the grain boundary, $S_n$
($S_p$) is the electron (hole) surface recombination velocity, $\bar
n\GB$ and $\bar p\GB$ are
\begin{align}
\label{nbar}
    \bar n\GB(E) &= N_C e^{\left(-E_g+E\right)/k_BT}\\
    \bar p\GB(E) &= N_V e^{-E/k_BT}
\label{pbar}
\end{align}
where $E$ is a defect energy level calculated from the valence band edge, $N_C$
($N_V$) is the conduction (valence) band effective density of states, $E_g$ is
the material bandgap, $k_B$ is the Boltzmann constant and $T$ is the
temperature.  The parameters $S_{n,p}^k$ and $\rho_k$ are related to the
electron and hole capture cross sections $\sigma_{n,p}^k$ by $S_{n,p}^k =
\sigma_{n,p}^k v_t \int_0^{E_g}\mathrm{d}E\ \rho_k(E)$, where $v_t$ is the thermal velocity. In this work we
vary $S_{n,p}^k$ with fixed $\rho_k$; this corresponds to varying
$\sigma_{n,p}^k$ accordingly.  Note that at thermal equilibrium \eq{eq:fGBe}
reduces to the Fermi-Dirac distribution, and is independent of the recombination
velocity parameters. In this case the occupancies of donor and acceptor states
at energy $E$ are equal, which we denote by $f\GB(E)$ where $f\GB(E) =
(1+\exp[(E-E_F)/k_BT])^{-1}$.

In this work we restrict the scope of Eq.~(\ref{QGB}) to two cases of interest
for positively charged grain boundaries. First, we focus on
a single donor state at $E\GB$: $\rho_D(E)=N\GB \delta(E-E\GB)$, where
$N\GB$ is a two-dimensional defect density (units of $\rm m^{-2}$). Denoting
$f_0=f\GB(E\GB)$, the grain boundary charge reads
\be
Q\GB = qN\GB (1-f_0).
\label{QGB1}
\ee
Our second case, a continuum of gap states, is based on the observation that a
wide variety of states populate the band gap of polycrystalline thin film
materials~\cite{Johnson2005, Krasikov2017}. In the absence of precise knowledge
of electronic defect structure for these materials, we consider
a continuum of uniform densities of acceptor and donor states
in the gap: $\rho_D(E)=N\GB/E_g$, $\rho_A(E) = \alpha N\GB/E_g$. $\alpha$ is the
ratio of  acceptor to donor densities of states. For this case we use
$E\GB$ to denote the neutral energy level of the distribution of states: the
grain boundary charge is
zero when the gap states are filled to this level~\cite{Rohderick}. If the Fermi
level is above (below) $E\GB$, the grain boundary core develops a negative
(positive) charge.  At thermal equilibrium, the continuum of gap states can
therefore be mapped onto an effective {\it single} defect level positioned at
energy $E\GB$. The defect charge is then
\be
Q\GB = qN\GB (1-2f_0)
\label{QGB2}
\ee
where $f_0$ is now the effective occupancy of the effective single
donor+acceptor defect state $E\GB$. This latter case was studied in detail in
Ref.~\onlinecite{Gaury2016GB}. This mapping from a continuum of donor+acceptor
states to a single donor+acceptor state is valid for values of $\alpha$ such
that the densities of donor and acceptor states remain commensurate~\footnote{We
will restrict $\alpha$ to the interval $[0.1, 4]$}. In the large defect
density limit (specified below) we find that $E\GB=E_g/(1+\alpha)$.

\bigskip
We consider large defect densities of states such that
$Q\GB/(qN\GB) \ll 1$.  We show in Appendix~\ref{pinning} that the critical
defect density for this condition to be satisfied is
\be
    N_{\rm GB}^{\rm crit} = \frac{1+e^3}{q} \sqrt{8\epsilon N_A(E\GB-E_F + 3k_BT)}.
    \label{r1}
\ee
for the single donor state, and
\be
    \rho_D^{\rm crit} = \frac{1}{q} \frac{\sqrt{8\epsilon N_A
    (E\GB-E_F-k_BT)}}{
    k_BT \ln\left(\frac{1+e^{(E_g-E\GB+k_BT)/k_BT}}{1+e^{(-E\GB+k_BT)/k_BT}} \right)
    -\alpha E_g }
    \label{r2}
\ee
for the continuum of acceptor and donor states. For material parameters typical
of CdTe, \eq{r1} is on the order of $10^{12}~\rm cm^{-2}$ for $E\GB \in [0.4~\rm
eV, 1.3~\rm eV]$, and \eq{r2} ranges from $4\times 10^{10}~\rm cm^{-2}\cdot
eV^{-1}$ to $6\times 10^{11}~\rm cm^{-2}\cdot eV^{-1}$ for $\alpha \in [0.1,
4]$.

\section{Equilibrium properties and assumptions away from equilibrium}
\label{assumptions}
\subsection{Equilibrium properties}
We first provide an expression for the equilibrium built-in potential $V_{\rm
GB}^0$ in terms of the defect
occupancy $f_0$.  In both single donor and continuum defect cases,
the absence of recombination at thermal equilibrium gives the equilibrium
carrier densities along the grain
boundary~\footnote{The general net capture rate for electrons at the grain
boundary core is $r = S_n(1-f_0)n\GB - S_nf_0 \bar n\GB(E\GB)$. At thermal
equilibrium
this rate is zero which gives \eq{neq}. The same reasoning applies for holes and
leads to \eq{peq}.}
\begin{align}
\label{neq}
    n_{\rm GB}^{\rm eq} &= \frac{f_0}{1-f_0}\bar n\GB(E\GB) \\
    p_{\rm GB}^{\rm eq} &= \frac{1-f_0}{f_0} \bar p\GB(E\GB)~.
\label{peq}
\end{align}
We take the energy reference at the valence band edge in the grain interior of the
$p$-type material, as shown in Fig.~\ref{system}(c). With this reference, the
distance between the Fermi level and the conduction band at the grain boundary
core is $E_g - E_F - qV_{\rm GB}^0$. This result determines another form of the equilibrium
electron density at the grain boundary core, $n_{\rm GB}^{\rm eq} = N_C\exp[(E_F
- E_g + qV_{\rm GB}^0)/k_BT]$.  Equating this form with \eq{neq} leads to the
equilibrium grain boundary built-in potential
\be
    qV_{\rm GB}^0 = E\GB - E_F + k_BT \ln\left(\frac{f_0}{1-f_0} \right).
\label{VGB}
\ee

We next consider the determination of the equilibrium defect state occupancy
$f_0$. The value of $f_0$ is determined by electroneutrality: the grain boundary
charge must be compensated by the charge of the surrounding depletion region.
The depletion region width surrounding the grain boundary in the $p$-type region
is $W\GB = \sqrt{2\epsilon V_{\rm GB}^0/(qN_A)}$ as shown in Fig.~\ref{system}
(the schematic neglects the modification of the grain boundary built-in
potential in the $pn$ junction depletion region).   For the donor case, the
electroneutrality requirement leads to the following equation for $f_0$:
\be
\frac{1}{8}\left(\frac{N\GB}{N_A L_D}\right)^2 \left(1-f_0\right)^2= \frac{E\GB
- E_F}{k_B T} + \ln\left(\frac{f_0}{1-f_0}\right),
\label{eq:f0}
\ee
where
$L_D=\sqrt{\epsilon k_B T/q^2 N_A}$.  In general \eq{eq:f0} must be solved
numerically for $f_0$.  Since no closed form of $f_0$ is available, we present
our results in terms of the variable $f_0$.  Note that as $N\GB\rightarrow
\infty$, $f_0\rightarrow 1$ and the built-in potential of \eq{VGB} diverges
logarithmically.  In practice, realistic values of $N\GB$ are well below this
diverging limit, so this issue can be safely ignored.  In the continuum defect
case, $f_0=1/2$ for the assumed large value of $N\GB$.  In this case,
\eq{VGB} reduces to the previously studied single donor+acceptor case of Ref.
\onlinecite{Gaury2016GB}.

%========================================================
\subsection{Assumptions in the nonequilibrium analysis}
%========================================================
A direct analytical solution for the full two-dimensional problem is not
feasible. To make analytical progress, we split the two-dimensional system into
two one-dimensional sub-systems: the grain boundary core where electrons are
electrostatically confined~\cite{gloeckler2005grain, Gaury2016GB, Tuteja2016},
and the grain interior (grain boundary free $pn$ junction) where a solution is
known.
Our approach relies on approximations (or assumptions) which
connect these two problems, and render the continuity-Poisson equations along
the grain boundary core analytically tractable.  In this
section we state our assumptions and sketch out the subsequent solution
procedures.

\bigskip
One blanket assumption is that the hole quasi-Fermi level is
approximately flat across and along the grain boundary. This is valid
because the hole current along the grain boundary is
negligible (electrons carry the current along the grain boundary), while in the
grain interior holes are majority carriers.  We provide a criterion restricting
the validity of this assumption in Appendix~\ref{flat}. We find that for typical
material parameters of CdTe, this assumption is generally valid for intragrain
hole mobilities on the order of $50~{\rm cm^2/(V\cdot s)}$.

The next assumption is that the grain boundary charge does not change with
voltage. This is justified by the limit of high defect density of states
$Q\GB(V)/(qN\GB)\ll 1$. This assumption is crucial as it enables us to relate the
electrostatic potential to the quasi-Fermi levels without the Poisson equation.
We denote the nonequilibrium defect occupancy with $f$, which replaces $f_0$ in
Eqs.~(\ref{QGB1}) and (\ref{QGB2}) for systems out of equilibrium.
$f$ is an integral of \eq{eq:fGBe} and depends on the
nonequilibrium carrier densities. Fixing the grain boundary charge to its
equilibrium value results in assuming $f=f_0$.  The relative sizes of the terms
in \eq{eq:fGBe} delineate three regimes of different behavior:
\begin{description}
\item [``$n$-type'' grain boundary]
In this case, the defect occupancy is determined by the electron carrier density
at the grain boundary.  $f$ remains fixed by maintaining a constant distance
between electron quasi-Fermi level and (actual or effective) $E\GB$.  We further
assume that the electron quasi-Fermi level is relatively flat and equal to its
bulk value. This is valid because the high electron density in the grain
boundary core enables high currents with relatively small quasi-Fermi level
gradients.  Together with the assumption of the relatively flat hole quasi-Fermi
level, the densities and recombination are easily determined.

\item  [``$p$-type'' grain boundary]
In this case, the occupancy of the defect state(s) is determined by the hole
carrier density at the grain boundary.  $f$ remains fixed by maintaining a
constant distance between the hole quasi-Fermi level and (actual or effective)
$E\GB$.  Because the electron density is small at the grain boundary core, the
electron quasi-Fermi level develops gradients to drive the electron current
along the grain boundary.  In this case we must solve a one-dimensional
diffusion equation for the electron density to obtain the carrier densities and
recombination.

\item  [High recombination]
For sufficiently large applied voltages, the electron and hole carrier densities
are the largest terms in the expression for $f$ and determine the defect
occupancy.  In the donor case, maintaining $f=f_0$ leads to the following
relation between electron and hole density:
\be
    S_pp\GB \approx (1-f_0) S_nn\GB.  \label{neqp1}
    \ee
In the continuum case, the occupancy of the acceptor and donor states are
constrained to ensure that $f\approx 1/2$. Because $f$ is now an integral of the
acceptor and donor occupancies, a simple relation like \eq{neqp1} does not
exist. We show in Sec.~\ref{pro} that $f=1/2$ leads to the relation:
\be
    p\GB = \gamma(V) n\GB
    \label{ga}
\ee
where the density ratio $\gamma$ varies weakly with voltage.
In this high-recombination regime, \eq{neqp1} or \eq{ga}
together with the assumption of flat hole quasi-Fermi level leads to a
one-dimensional drift-diffusion equation for electrons confined to the grain
boundary.  Solving this equation leads to the carrier densities and
recombination.
\end{description}

A final assumption which applies for all of our analysis is that the depletion
regions of grain boundaries do not overlap, i.e. grain sizes $d$ are greater than
$2W\GB$. In other words, we assume that the electrostatic potential
of the grain interior is the same as that of a grain boundary free $pn$ junction.
This assumption is necessary because we need {\it a priori} knowledge of the
solution in the bulk in order to construct the solution along the grain
boundary.
For a doping density $10^{15}~\rm{cm^{-3}}$ this requirement implies
$d> 2~\mu \rm{ m}$.  The average grain size in CdTe thin films (excluding twin
boundaries) was recently~\cite{Moseley2015} found to be $2.3~\mu \rm m$.

%=================================
\section{Grain boundary dark current of a single donor defect state}
%=================================
\label{donor}
We begin with the case of a single donor state in the gap of the absorber material.
The grain boundary charge is proportional to $1-f_0$ (see \eq{QGB1}).  In the limit of
large defect density of states, the electroneutrality of the grain boundary is
satisfied when the defect level is fully occupied: $f_0 \approx 1$. In the
$n$-region ($x<x_0$) the large concentration of electrons satisfies this
requirement without the need for modifying the electrostatics around the grain
boundary.  In the $p$-region, however, the resulting built-in potential around
the grain boundary is given by \eq{VGB}
\be
    qV_{\rm GB}^0 = E\GB - E_F - k_BT \ln(1-f_0).
    \label{vgbD}
\ee
Because of the logarithm term and $f_0 \approx 1$, the Fermi level $E_F$ is not
pinned to $E\GB$.

\begin{figure}
 \includegraphics[width=0.49\textwidth]{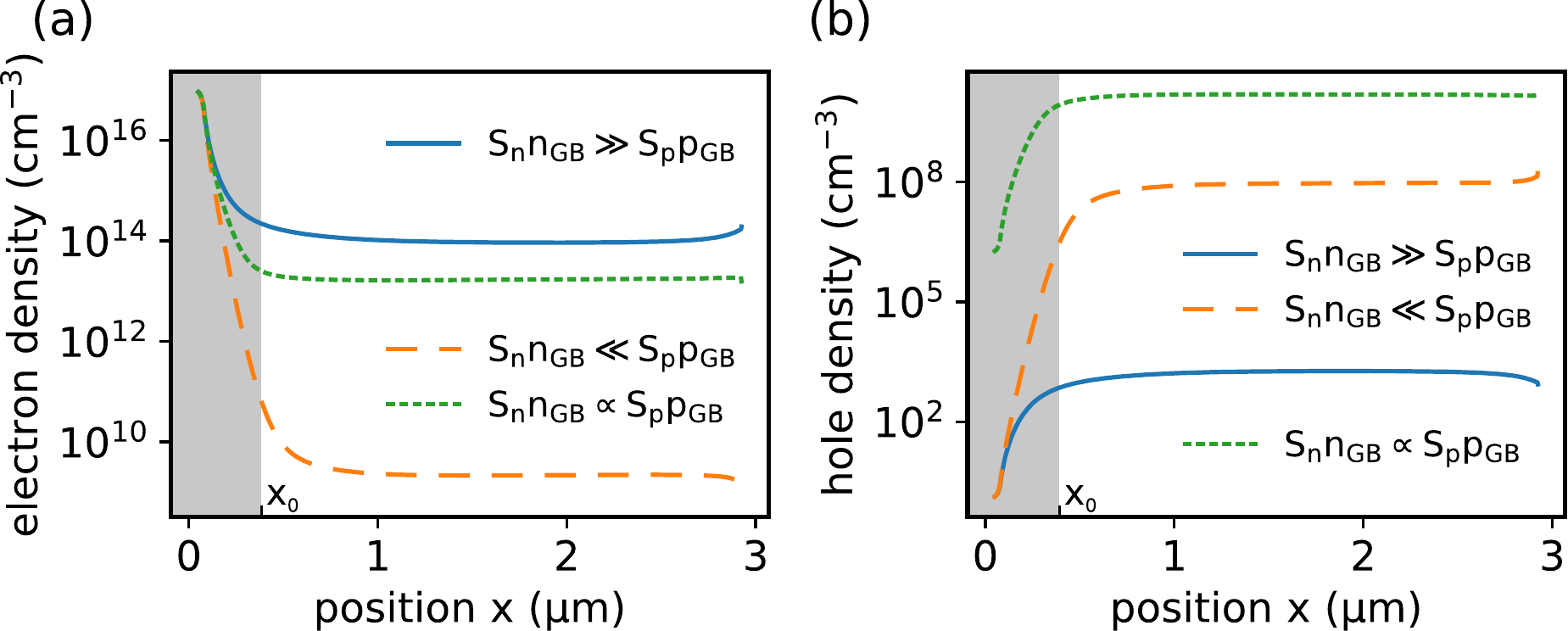}
 \caption{\label{2plots} Numerical simulation results for carrier densities
 along the grain boundary for the three regimes determined by the ratio
 $S_nn\GB$ to $S_p\bar p\GB$.  (a)
 Electron density. (b) Hole density. $S_n n\GB \gg S_p\bar p\GB$ was obtained for
 $E\GB=1~\rm eV$ at $V=0.3~\rm V$ (blue continuous lines), $S_nn\GB \ll
 S_p\bar p\GB$ for $E\GB=0.5~\rm eV$ at $V=0.3~\rm V$ (orange dashed lines) and
 $S_nn\GB \propto S_pp\GB$ for $E\GB=0.5~\rm eV$ at $V=0.7~\rm V$ (green
 dotted lines). All calculations were done for $S_n=S_p=10^5~\rm{cm/s}$,
 $\mu_n=320~\rm{cm^2/(V\cdot s)}$, $\mu_p=40~\rm{cm^2/(V\cdot s)}$ and
 $N_A=10^{15}~\rm{cm^{-3}}$.  General parameters are listed in
 Table~\ref{params}.  }
\end{figure}

\bigskip
We now derive analytical expressions for the dark recombination current at the
grain boundary. We support the physical descriptions with numerically computed
carrier densities along the grain boundary presented in Fig.~\ref{2plots}. The
absence of modification of the electrostatics in the $n$-region results in a
recombination similar to the grain interior of the $pn$ junction depletion
region: the recombination is determined by the hole density (minority carrier
density) which decreases sharply for $x<x_0$, as shown in Fig.~\ref{2plots}(b).
In what follows we therefore neglect the contribution of this part of the grain
boundary to the total recombination, and focus on carrier densities and
recombination for $x\geq x_0$. The general expression for the grain boundary
recombination current reads
\be
    J\GB(V) = \frac{1}{d} \int_{x_0}^{L\GB} \int_0^d \mathrm{d}x\mathrm{d}y\
    R\GB(x,y),
    \label{JGB}
\ee
where $L\GB$ is the length of the grain boundary. $R\GB$ is the recombination at
the grain boundary and has the Schockley-Read-Hall form
\be
    R\GB(x,y) = \frac{S_nS_p (n\GB(x) p\GB(x) - n_i^2)\delta(y-y\GB)}{S_n(n\GB(x)+\bar n\GB) +
    S_p(p\GB(x)+\bar
    p\GB)},
    \label{RGB1}
\ee
where $n_i$ is the intrinsic carrier density and we dropped the ``donor''
superscript for the recombination velocities. $\bar n\GB$ and $\bar p\GB$ are
given by Eqs.~(\ref{nbar}) and (\ref{pbar}) evaluated at $E=E\GB$.

As discussed in Sec.~\ref{assumptions}, under nonequilibrium conditions we
assume that the grain boundary carrier densities evolve while keeping the level
occupancy equal to its thermal equilibrium value $f_0$.  Using this
assumption and comparing the relative sizes of the terms in the nonequilibrium
level occupancy \eq{eq:fGBe} leads to three
regimes of interest for the grain boundary dark current. We next analyze these
regimes individually.

%=============================================================================
\subsection{Grain boundary recombination for $S_pp\GB \ll S_n \bar n\GB$}
%=============================================================================
\label{ntype}

We first consider $S_pp\GB \ll S_n\bar n\GB$, also called ``$n$-type'' regime.
As discussed in Sec.~\ref{assumptions}, for this case $f_0$ remains fixed by
keeping $E_{F_n}-E\GB$ constant. Equivalently, $E_{F_n}$ replaces $E_F$ in
\eq{vgbD}\cite{Gaury2016GB}. In the grain interior of the $p$-region,
the increase of voltage $V$ shifts the electron quasi-Fermi level from the
valence band by an amount $qV$.  The electron current transverse to the grain
boundary is small compared to the longitudinal one. So despite the low electron
density in the grain interior, the gradient of $E_{F_n}$ across the grain
boundary driving the transverse current is small and can be neglected. Assuming
that $E_{F_n}$ is flat across the grain boundary, the built-in potential also
varies with $V$:
\begin{align}
    qV\GB &= E\GB - E_{F_n} - k_BT\ln(1-f_0) \nonumber\\
         &= E\GB - E_{F} - qV - k_BT\ln(1-f_0) \nonumber\\
         &= q(V_{\rm GB}^0 - V),
    \label{vgb}
\end{align}
where $E_F$ is the equilibrium Fermi level.
Equation~(\ref{vgb}) shows that the grain boundary built-in potential decreases
linearly with voltage for $x>x_0$. This is shown in Fig.~\ref{donor_ntype}(b).
The reduction of the barrier allows holes of the grain interior to flow toward
the grain boundary core. The recombination of
holes generates an electron current along the grain boundary, as depicted in
Fig.~\ref{donor_ntype}(a).
 \begin{figure}
 \includegraphics[width=0.49\textwidth]{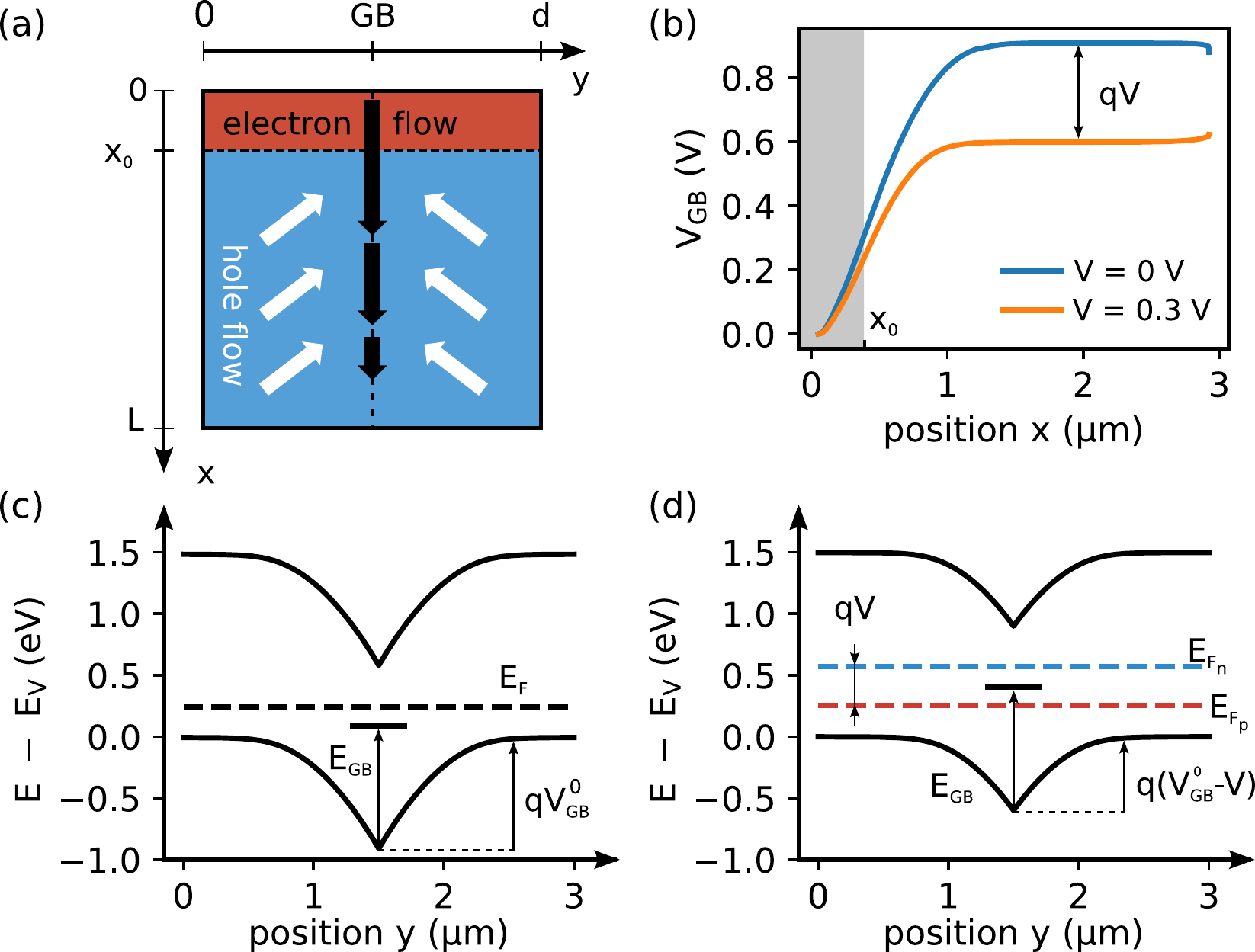}
\caption{\label{donor_ntype} (a) Schematic of the electron and hole currents in
the regime $S_pp\GB \ll S_n\bar n\GB$. (b) Difference in electrostatic potential
between grain boundary and grain interior $V\GB$ along the grain boundary for
the applied voltages $V=0~\rm V$ and $V=0.3~\rm V$. (c) and (d)
Band diagrams across the grain boundary for $x>x_0$, for $V=0~\rm V$ and
$V=0.3~\rm V$ respectively.}
\end{figure}

Because holes are majority carriers in the bulk of the
absorber, the hole quasi-Fermi level is flat and equal to $E_F$. We derived
in Appendix~\ref{flat} a criterion under which the hole quasi-Fermi level is
flat across the grain boundary, so that the bulk quasi-Fermi level extends to
the grain boundary core.
Using \eq{vgb} and the assumption of flat hole quasi-Fermi level, the distance
between $E_{F_p}$ and the valence band is $E\GB-qV - k_BT\ln(1-f_0)$.
The hole density at the grain boundary therefore reads
\be
    p\GB(x>x_0) = (1-f_0) N_V e^{(-E\GB +qV)/k_BT}.
    \label{assumption_efp_flat}
\ee
The hole density is shown in Fig.~\ref{2plots}(b) (blue continuous
lines). The electron density is equal to its equilibrium value \eq{neq} for
$x>x_0$, shown in Fig.~\ref{2plots}(a).
Because $S_n n\GB$ is much larger than all the other terms in the denominator of
\eq{RGB1}, the recombination simplifies to
\be
    R\GB (x>x_0) = S_p p\GB(x>x_0)
\ee
for $V\gg k_BT/q$. The recombination is uniform for $x>x_0$ and negligible in
the $n$-region, so the dark recombination current reads
\be
    \label{j1}
    J\GB(V) = (1-f_0) \frac{S_p(L\GB-x_0)}{d} N_V e^{(-E\GB+qV)/k_BT}.
\ee
The features of \eq{j1} are a saturation current $(1-f_0)S_p(L\GB-x_0)/d$, an ideality
factor of 1, and an activation energy $E\GB$.

\bigskip
In Appendix~\ref{flat} we derive a condition under which the hole quasi-Fermi
level is approximately flat. This condition reads ($V_T=k_BT/q$)
\be
    \frac{S_p}{2\mu_p}\sqrt{\frac{2\epsilon}{qV_TN_A}} < 1.
    \label{eflat}
\ee
For $S_{n,p}=10^5~\rm cm/s$, $N_A=10^{15}~\rm cm^{-3}$,
$\epsilon=9.4~\epsilon_0$, $V_T=25~\rm meV$, \eq{eflat} is satisfied for
$\mu_p>32~\rm cm^2/(V\cdot s)$. Over the course of this work, we found this
numerical value to be an acceptable threshold for all the grain boundary regimes of
this section.

%=============================================================================
\subsection{Grain boundary recombination for $S_nn\GB \ll S_p\bar p\GB$}
%=============================================================================
We turn to $S_nn\GB \ll S_p\bar p\GB$, also called ``$p$-type'' regime.
In this case the distance between $E\GB$ and
$E_{F_p}$ does not change with the applied voltage $V$, as seen in
Figs.~\ref{donor_ptype}(c) and (d).
We further assume that $E_{F_p}$ is  flat equal to
$E_F$~\footnotemark[\value{footnote}]. As a result \eq{vgbD}
shows that the grain boundary
built-in potential is independent of the applied voltage for $x>x_0$. This is
shown in Fig.~\ref{donor_ptype}(b).
\begin{figure}
 \includegraphics[width=0.49\textwidth]{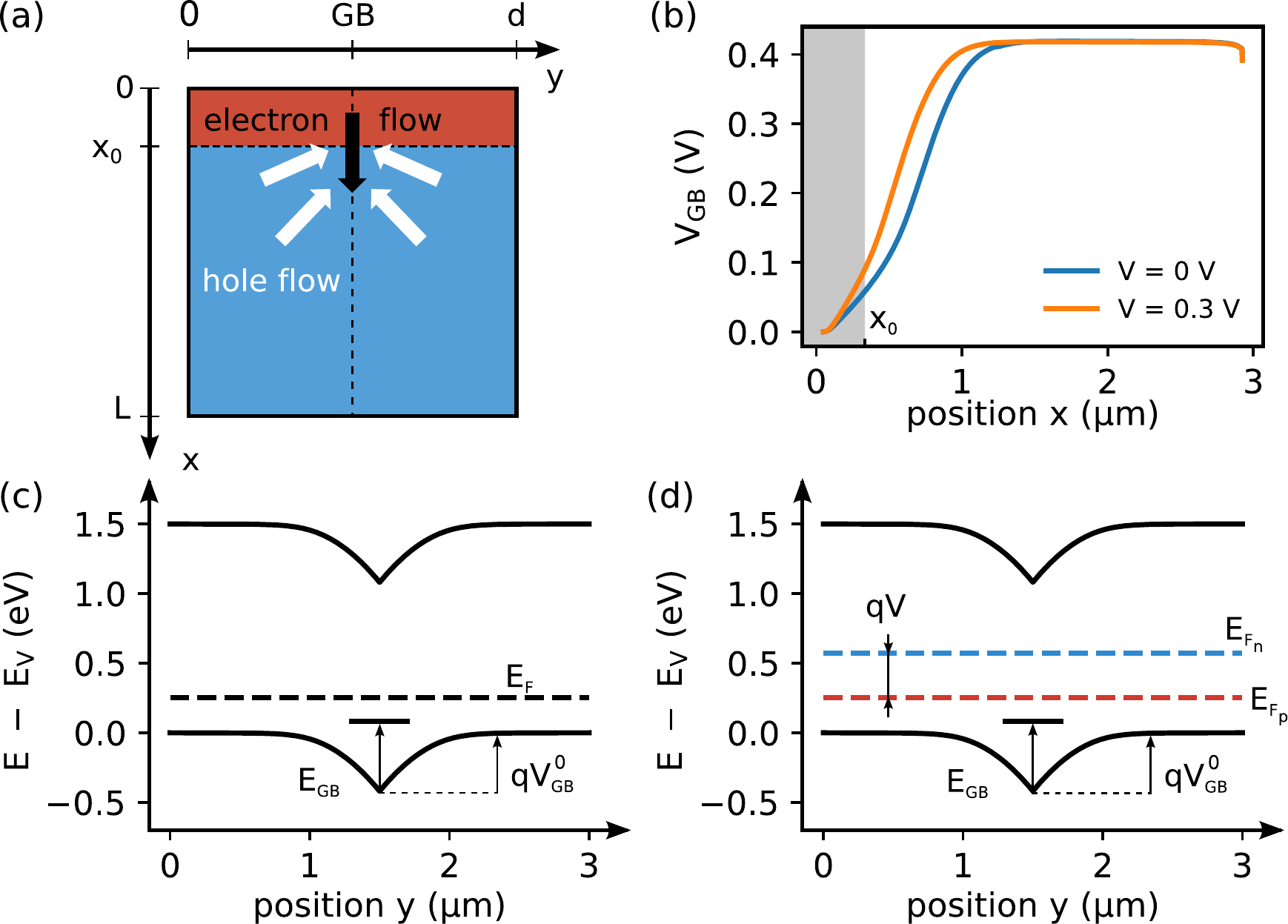}
\caption{\label{donor_ptype} (a) Schematic of the electron and hole currents in
the regime $S_nn\GB \ll S_p\bar p\GB$. (b) Difference in electrostatic potential
between grain boundary and grain interior $V\GB$ along the grain boundary for
the applied voltages $V=0~\rm V$ and $V=0.3~\rm V$. (c) and (d)
Band diagrams across the grain boundary for $x>x_0$, for $V=0~\rm V$ and
$V=0.3~\rm V$ respectively.}
\end{figure}

\bigskip
The electron transport is more complex than in the previous case. In particular,
the electron quasi-Fermi level is not always equal to $E_F+qV$ along the grain
boundary, but varies significantly to accommodate the electron current. The grain boundary
built-in potential confines electrons near the grain boundary core, leading to a
one-dimensional motion along it. As a result, the continuity equation around the
grain boundary reduces to a one-dimensional equation along the grain boundary
core ($x$-direction).  Upon solving this equation beyond $x_0$ (see
Appendix~\ref{derivations:nsmall}), the electron density reads
\be
    n\GB(x>x_0) = \frac{N_C}{1-f_0} e^{(-E_g+E\GB + qV)/k_BT}
    e^{-\frac{x-x_0}{L_n}},
\ee
where $L_n = \sqrt{2D_nL_{\mathcal{E}}/(S_n (1-f_0))}$ ($D_n=k_BT\mu_n/q$:
electron diffusion coefficient) is the diffusion length of electrons along the
grain boundary. $L_{\mathcal{E}}$ is the length scale of the confinement of the
electrons to the grain boundary core. This length is related to $\mathcal{E}_y$,
the electric field transverse to the grain boundary in the neutral region of the
$pn$ junction, by $2L_{\mathcal{E}} = 2V_T/\mathcal{E}_y$ (see \eq{ey} in
Appendix~\ref{derivations:nsmall}).
The electron density described here is shown in
Fig.~\ref{2plots}(a) (dashed orange line). The hole density (shown in
Fig.~\ref{2plots}(b)) is equal to its equilibrium value \eq{peq}.
Because $S_p\bar p\GB$ dominates the denominator of \eq{RGB1}, the recombination reads
\be
    R\GB(x>x_0) = (1-f_0) S_n  n\GB(x>x_0)
\ee
for $V\gg k_BT/q$. We now consider two limiting cases for the integration of the
recombination.

The first limit is the large diffusion length, $L_n\gg L\GB$, obtained for large
mobilities and small values of recombination velocity. The electron density is
uniform for $x>x_0$, leading to the dark recombination current
\be
    J\GB(V) = \frac{S_n (L\GB-x_0)}{d} N_C e^{(-E_g+E\GB +qV)/k_BT}.
    \label{j1pp}
\ee
In the second limit, $L_n\ll L\GB$, the electron mobility is small so that the
electron density decays rapidly beyond $x_0$. The recombination is
peaked on both sides of $x_0$ and the recombination current reads
\be
    J\GB(V) = \frac{S_n L_n}{d}  N_C e^{(-E_g+E\GB +qV)/k_BT}.
    \label{j1p}
\ee
The description of this regime is shown in Fig.~\ref{donor_ptype}(a). Holes
converge to $x_0$ where they recombine, generating a localized electron current.
Both regimes have similar features: the saturation current varies as $S_nN_C/d$,
the ideality factor is 1 and the thermal activation energy is $E_g-E\GB$.

%=============================================================================
\subsection{Grain boundary recombination for $S_nn\GB \approx (1-f_0) S_pp\GB$}
%=============================================================================
\label{n_prop_p}
As we increase the applied voltage in either of the previous cases, $S_n \bar
n\GB$ and $S_p \bar p\GB$ become negligible compared to $S_pp\GB$ and $S_nn\GB$
respectively.  Because the system approximately maintains the level occupancy
close to its thermal equilibrium value, the carrier densities satisfy the
relation
\be
    f_0 \approx \frac{S_n n\GB}{S_nn\GB + S_pp\GB}.
    \label{tu}
\ee
Equation~(\ref{tu}) leads to $S_pp\GB \approx (1-f_0) S_nn\GB$, defining the
``high-recombination'' regime.

While the electrostatic potential now varies along the grain boundary, the
built-in potential still confines the electrons to one-dimensional motion
along the grain boundary core. Similarly to the regime $S_nn\GB \ll S_p\bar
p\GB$, a one-dimensional continuity equation describes the electron transport
along the grain boundary. Upon solving this equation
(see Appendix~\ref{efndemo}), we find the carrier densities
\begin{align}
    \label{n'}
    n\GB(x>x_0) &= \frac{1}{\sqrt{1-f_0}} \sqrt{\frac{S_p}{S_n}}n_i
    e^{qV/(2k_BT)} e^{-\frac{x-x_0}{L_n'}}\\
    p\GB(x>x_0) &= \sqrt{1-f_0} \sqrt{\frac{S_n}{S_p}}n_i e^{qV/(2k_BT)}
    e^{-\frac{x-x_0}{L_n'}},
    \label{px}
\end{align}
where $L_n' = \sqrt{4D_nL'_{\mathcal{E}}/(S_n (1-f_0))}$ ($L'_{\mathcal{E}}$ is
the characteristic length of the electric field across the grain boundary).
These densities yield the grain boundary recombination
\be
    R\GB(x>x_0) = \sqrt{1-f_0} \sqrt{S_nS_p} n_i e^{qV/(2k_BT)} e^{-\frac{x-x_0}{L_n'}}
    \label{Rd}
\ee
for $V\gg k_BT/q$. We consider two limiting cases for the recombination
current.

Because of the factor $1-f_0$ in $L_n'$, common material parameters for CdTe
lead to large diffusion lengths such that
$L_n'\gg L\GB$. The uniform grain boundary (for $x>x_0$) described in
Sec.~\ref{ntype} applies in this case, and the recombination current reads
\be
    J\GB(V) = \sqrt{1-f_0} \frac{\sqrt{S_nS_p} (L\GB-x_0)}{d} n_i
    e^{qV/(2k_BT)}.
    \label{lim1}
\ee
For smaller values of the electron diffusion length, $L_n'\ll L\GB$
and the recombination is peaked at $x_0$ with the electron and hole flows
depicted in Fig.~\ref{donor_ptype}(a). The dark recombination current
therefore reads
\be
    J\GB(V) = \sqrt{1-f_0} \frac{\sqrt{S_nS_p} L_n'}{d} n_i e^{qV/(2k_BT)}.
    \label{lim2}
\ee
In both limits the thermal activation energy is $E_g/2$ and the ideality factor
is 1. Note that the factor $1-f_0$ is on the order of $10^{-4}$ to $10^{-3}$ for
typical doping densities. As a result, $S_n$ is effectively reduced by two
orders of magnitude, significantly reducing the amplitude of the grain
boundary recombination current. 

For equal $E\GB$, this difference between the single donor case
and the single acceptor+donor state studied in Ref.~\onlinecite{Gaury2016GB}
comes from the difference in band bending associated with the two cases.
For the donor case, the band bending is substantially
increased (see Eq.~(\ref{vgbD})) relatively to the donor+acceptor case.  The
increased band bending leads to suppressed hole density, which in turn
suppresses electron-hole recombination.

%===============================================================
\subsection{Numerical verification of the analytical results}
%===============================================================
\label{3D}
We verify the accuracy of our analytical results using numerical solutions of
the drift-diffusion-Poisson equations. We used our own finite-difference
software to solve these equations for our geometry in
Fig.~\ref{system}(a).  We discretized the electron/hole currents using the
Scharfetter-Gummel scheme \cite{gummel1964self}, and used the Newton-Raphson
method to find the self-consistent solution.  To determine the electrostatic
potential boundary conditions, we performed two steps. First, we solved the
thermal equilibrium Poisson equation with $\partial \phi/\partial x=0$ at each
contact. Then, we solved the full drift-diffusion-Poisson equations by imposing
$\phi(x=0,y) = \phi^{eq}(x=0,y)+qV$ and $\phi(x=L,y)=\phi^{eq}(x=L,y)$, where
$\phi^{eq}$ is the equilibrium potential.   Table~\ref{params} gives the list of
material parameters used for these calculations.
\begin{table}
\begin{tabular}{lclc}
  \toprule
  Param. & Value & Param. & Value \\ \midrule
  $L$ & $3~ \mu{\rm m}$ &  $N_D$ & $10^{17}~{\rm cm^{-3}}$\\
  $d$ & $3~ \mu{\rm m}$  & $N_A$ & $(3\times 10^{14}~\mathrm{to}~10^{16})~{\rm cm^{-3}}$ \\
  $N_C$ & $8\times10^{17}~{\rm cm^{-3}}$ & $\tau_{n,p}$ & $(10~\mathrm{to}~100)~{\rm ns}$ \\
  $N_V$ & $1.8\times10^{19}~{\rm cm^{-3}}$ & $S_{n,p}$ & $(1~\mathrm{to}~10^6)~{\rm cm/s}$ \\
  $E_g$ & $1.5~{\rm eV}$   & $\mu_p$ & $40~\rm{cm^2/(V\cdot s)}$ \\
  
  $\epsilon$ & $9.4~\epsilon_0$ & $\mu_{n}$ & $(10~\mathrm{to}~10^3)~{\rm
  cm^2/\left(V\cdot s\right)}$ \\
  $N\GB$ & $10^{14} ~{\rm cm^{-2}}$ \\
  \bottomrule\\
\end{tabular}
\caption{List of default parameters (param.) for numerical simulations.
Minority carrier lifetimes correspond to the lower range found in single
crystal CdTe~\cite{Kuciauskas2015, Burst2016}. Mobilities are varied across a
wide range of literature values~\cite{Suzuki2002, Sellin2005, Fink2006,
Duenow2014}. Lifetimes and surface recombination velocities are
taken equal for electrons and holes. \label{params} }
\end{table}

\bigskip
Figure~\ref{4plots_D} presents calculations for the grain boundary dark
current. At each applied voltage the current is given by the smallest value
between the $n$-type, $p$-type and high recombination regimes.
\begin{figure}
   \includegraphics[width=0.49\textwidth]{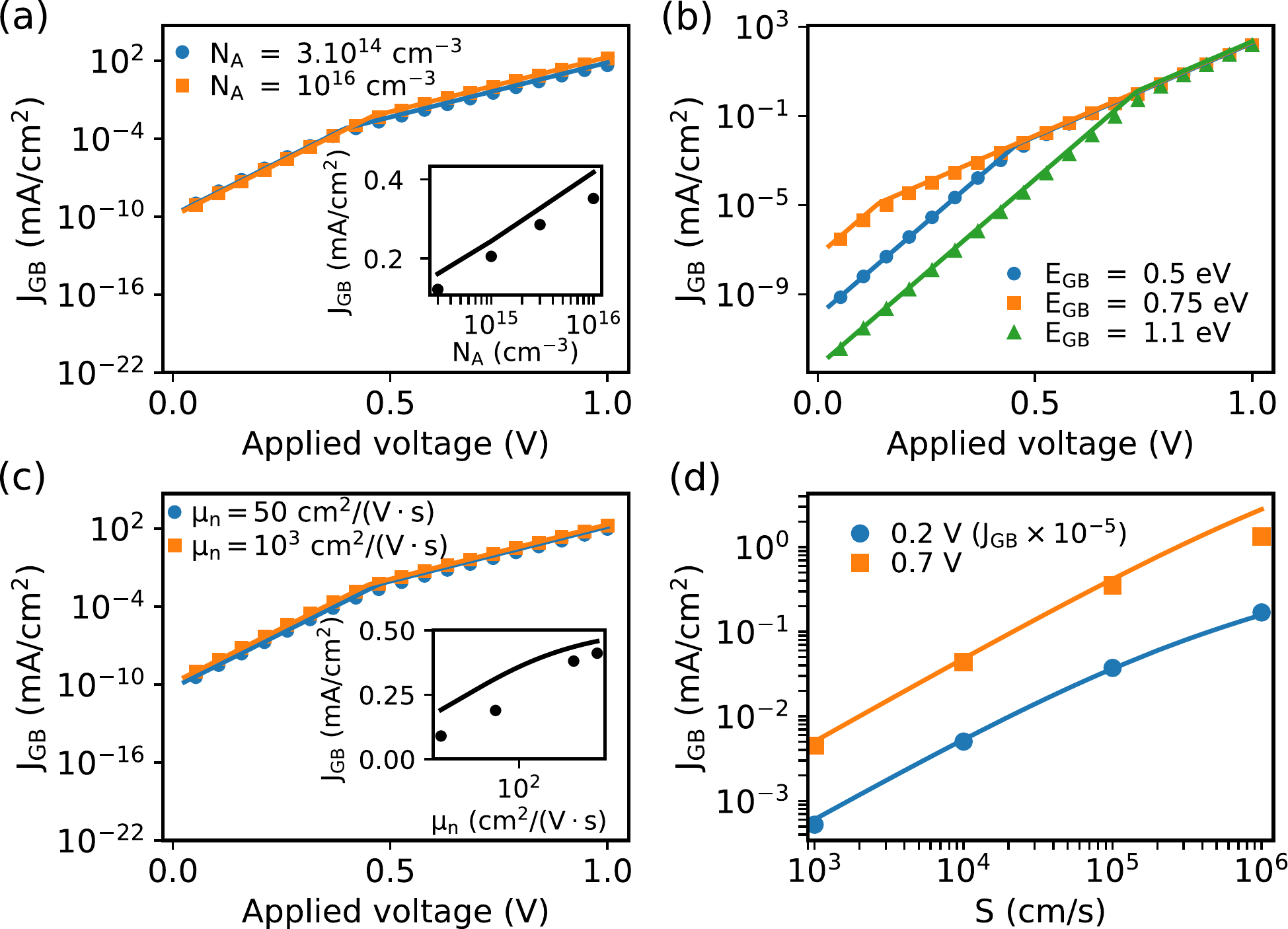}
    \caption{\label{4plots_D} Grain boundary recombination current characteristics
    $J\GB(V)$ for a single donor state at $E\GB=0.5~\rm{eV}$ ($p$-type grain
    boundary), $N_A=10^{16}~\rm{cm^{-3}}$, 
    $S_n=S_p=10^5~\rm{cm/s}$ and
    $\mu_n=320~\rm{cm^2/(V\cdot s)}$ unless specified otherwise. Symbols are
    numerical calculations, full lines correspond to analytical results
    Eqs.~(\ref{j1}), (\ref{JGBnsmall}) and (\ref{jgbn=p}). (a)
    $J\GB(V)$ varied with doping density. Inset: grain boundary recombination
    current as a function of doping density at $V=0.7~\rm{V}$. (b)
    $J\GB(V)$ varied with defect energy level. (c)
    $J\GB(V)$ varied with electron mobility. Inset:  grain boundary
    recombination current as a function of electron mobility for $V=0.7~\rm{V}$
    (high recombination regime). (d) Grain boundary
    recombination current as a function of surface recombination velocity
    ($S_n=S_p$), at $V=0.2~\rm{V}$ (dots) and $V=0.7~\rm{V}$ (squares).}
\end{figure}

The doping density is varied in Fig.~\ref{4plots_D}(a) for $E\GB=0.5~\rm eV$.
The crossover from the $p$-type regime ($S_n n\GB \ll S_p \bar p\GB$) to the
high recombination regime ($S_n n\GB \approx (1-f_0) S_p p\GB$) occurs at about
$0.5~\rm V$, as seen from the slope change. The inset shows the doping
dependence of the high recombination regime.  Contrary to the case studied in
Ref.~\onlinecite{Gaury2016GB} where common material parameters for CdTe lead to
decreasing grain boundary dark currents with doping at high voltages, the donor
state shows the opposite behavior. The key difference between these cases lies
in the effective electron surface recombination velocity entering the definition
of the electron diffusion length at high voltages. In the single acceptor+donor
case it is given by $S_n/2$, while in the single donor case we found
$(1-f_0)S_n$. Because $f_0\approx 1$ the latter value is orders of magnitude
smaller than the former. The primary consequence is that for the same $S_n$, the
diffusion length is much larger in the donor case. In turn, the limit $L_n' \gg
L\GB$ is the relevant one for common CdTe parameters with the single donor state
(\eq{lim1}), while the opposite limit is relevant for the case studied in
Ref.~\onlinecite{Gaury2016GB}. An increasing doping density increases the
depletion charge, leading to a larger value of $1-f_0$. Consequently the $1-f_0$
factor in \eq{lim1} is responsible for the increase in recombination current
with doping density observed here.  This increase is a major difference that
will reflect in the open-circuit voltage.

We show the various grain boundary types in Fig.~\ref{4plots_D}(b), where
$E\GB=0.5~\rm eV$ and $E\GB=1.1~\rm eV$ correspond respectively to a $p$-type and $n$-type
grain boundary at equilibrium. Figure~\ref{4plots_D}(c) shows the dependence of
the grain boundary dark current with electron mobility. This dependence is absent at low
voltage, as
shown by the limiting case \eq{j1pp}. At higher voltages, the relevant limit is
\eq{lim1}, as discussed above, which is independent of mobility. This limit
explains the weak mobility dependence shown in the inset.
Finally, Fig.~\ref{4plots_D}(d) shows the scalings of the grain boundary dark
current with surface recombination velocity.
We show grain boundary dark currents in the $p$-type regime
(dots) and high-recombination regime (squares). On this plot both scalings are
equal. The dark current scales as $S_p$ at low voltage, and as $\sqrt{S_nS_p}$
at high voltage. The second scaling is given by \eq{lim1} which is the relevant
limit in this case. With $S_n=S_p$ these scalings are identical.

%========================================================================
%========================================================================
\section{Grain boundary dark recombination current of a continuum of defect
states}
%========================================================================
%========================================================================
\label{continuum}
We turn to the case of a continuum of donor and acceptor states in the absorber band gap. We
assume densities of states uniform in energy: $\rho_D=N\GB/E_g$,
$\rho_A(E)=\alpha N\GB/E_g$, where
\be
    \alpha = \rho_A / \rho_D
\ee
determines the ratio of acceptor to donor densities of states. More
acceptor (donor) states leads to $p$-type ($n$-type) grain boundary core. The
neutral energy level of the distribution of gap states is $E\GB=E_g/(1+\alpha)$ (see
Sec.~\ref{model}).

\bigskip
Under nonequilibrium conditions the grain boundary dark current is given
by \eq{JGB} where the integral along the grain boundary ($x$-direction) now
starts from $x=0$. The recombination is the sum of the contributions from the
acceptor and donor states (represented by the superscript $k\in (A,D)$)
\begin{align}
    &R_{\rm GB}^k(x) =\nonumber\\
    &\int_0^{E_g}\frac{\mathrm{d}E}{E_g}\ \frac{S^k_n S^k_p(n\GB p\GB -
    n_i^2)}{S^k_n(n\GB+\bar n\GB(E)) + S^k_p(p\GB + \bar p\GB(E))}.
    \label{RGB}
\end{align}
Despite
the apparent complexity of a continuum of states as opposed to a single state,
the physical description of the nonequilibrium electron and hole currents is the
same as given in Ref.~\onlinecite{Gaury2016GB}.
This apparent complexity will be incorporated in
effective surface recombination velocities in what follows.
The results of Ref.~\onlinecite{Gaury2016GB} and of this section are gathered in
the last two rows of Table~\ref{form}.

%================================================================
\subsection{Grain boundary recombination for $S_n^k n\GB \gg S_p^k p\GB$}
%================================================================
We begin with the regime $S_n^k n\GB \gg S_p^kp\GB$ ($n$-type grain boundary). In
this regime the electron quasi-Fermi level is pinned to $E_g/(1+\alpha)$.
The recombination is determined by holes, which flow from the $p$-type grain interior
into the grain boundary core. Because most of the absorber
is $p$-type, the recombination is uniform along the entire grain boundary.
We refer to Sec.~III A of Ref.~\onlinecite{Gaury2016GB} for a more complete
description of this regime.

The electron density is independent of voltage and spatially uniform, given by \eq{neq} with
$f_0=1/2$. The hole density is also uniform and reads
\be
    p\GB = N_V e^{(-E\GB + qV)/k_BT}.
\ee
The grain boundary dark current reads
\be
    J\GB(V) = \frac{\mathcal{S}_p L\GB }{d} N_Ve^{\left(-E\GB+qV\right)/k_BT}
    \label{jc1}
\ee
where $\mathcal{S}_p$ is the effective surface recombination velocity
\be
    \mathcal{S}_p = \sum_{k \in {A,D}}S^k_p \int_{0}^{E_g}
    \frac{\mathrm{d}E}{E_g}\ \frac{1}{1+ \frac{\bar
    n\GB(E)}{n\GB} + \frac{S^k_p \bar p\GB(E)}{S^k_n n\GB}}.
    \label{Sp}
\ee
The integrals in \eq{Sp} can
be computed analytically but the results are cumbersome and difficult to interpret.
To give a sense of these integrals, we refer to the solid blue line of
Fig.~\ref{occupancy}(b) where we show the grain boundary recombination as a
function of energy. This figure
shows that only states with energies $E_g - E\GB \lesssim
E \lesssim E\GB$ contribute significantly to the recombination.
The upper limit results from the fact that states above $E\GB$ are empty of
electrons. The lower limit is the energy at which holes are emitted from the
defect state to the valence band faster than electrons relax from the conduction
band to the defect state. This rapid hole emission rate prevents recombination.
The typical width of the integrand of $\mathcal{S}_p$ is
\be
    \Delta_k(\alpha) = \left| \frac{1-\alpha}{1+\alpha}  + \frac{k_BT}{E_g}\ln\left(
    \frac{S_n^k}{S_p^k} \frac{N_C}{N_V} \right) \right|
    \label{delta}
\ee
as shown on the continuous plot of Fig.~\ref{occupancy}(b). Equation~(\ref{Sp})
therefore simplifies to
\be
    \mathcal{S}_p \approx S^A_p \Delta_A(\alpha) + S^D_p \Delta_D(\alpha).
\ee
Note that the occupancy of the gap
states is independent of the applied voltage in this regime (see the
continuous curve in Fig.~\ref{occupancy}(a)). That is because the occupancy of
the gap states is determined solely by the electron density, which is
independent of voltage here.

\bigskip
In Appendix~\ref{flat} we derive a condition under which the hole quasi-Fermi
level is approximately flat. This condition applied here reads ($V_T=k_BT/q$)
\be
    \frac{\mathcal{S}_p}{2\mu_p}\sqrt{\frac{2\epsilon}{qV_TN_A}} < 1.
    \label{efplat}
\ee
For $\alpha=2$, $S^{A,D}_{n,p}=10^5~\rm cm/s$, $N_A=10^{15}~\rm cm^{-3}$,
$\epsilon=9.4~\epsilon_0$, $V_T=25~\rm meV$, \eq{efplat} is satisfied for
$\mu_p>25~\rm cm^2/(V\cdot s)$. This value is in the range of standard bulk
mobilities for CdTe~\cite{Sellin2005, Duenow2014}.  We found this order of
magnitude to be an acceptable threshold for all the grain boundary regimes with
the continuum of defect states.
\begin{figure}
 \includegraphics[width=0.49\textwidth]{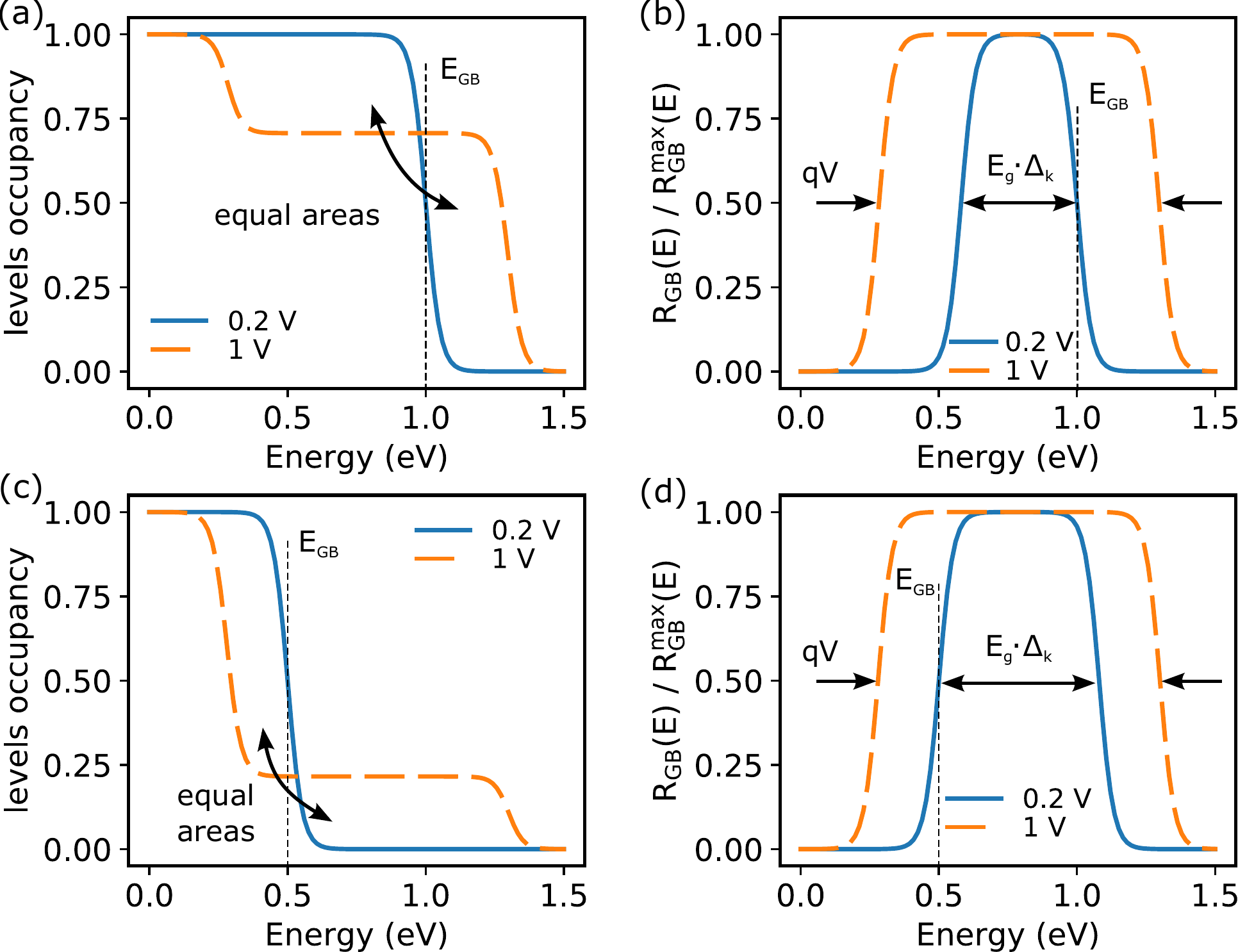}
\caption{\label{occupancy} (a), (c) Occupancy of the gap states as a function of
energy at $x_0$ for $\rho_A/\rho_D=0.5$ ($E\GB=1~\rm eV$) and $\rho_A/\rho_D=2$
($E\GB=0.5~\rm eV$) respectively. (b), (d)
Normalized recombination at $x_0$ as a function of energy corresponding to the
occupancy levels in (a) and (c) respectively. We used $S_{n,p}^A = S_{n,p}^D$ so
that recombinations of the acceptor and donor states are the same.}
\end{figure}

%================================================================
\subsection{Grain boundary recombination for $S_n^kn\GB \ll S_p^kp\GB$}
%================================================================
In the regime $S_n^kn\GB \ll S_p^kp\GB$ ($p$-type grain boundary), the hole
quasi-Fermi level is pinned to $E_g/(1+\alpha)$. In this regime the
recombination is determined by electrons flowing into the grain boundary core
from regions of the grain interior where $n > p$. These regions correspond to $x
< x_0$ in Fig.~\ref{system}(a). The recombination is therefore mainly
concentrated within the $n$- region of the $pn$ junction depletion region, and
is uniform for
$x < x_0$. We refer to Sec.~III B of Ref.~\onlinecite{Gaury2016GB} for a
complete description of this case.

For $x>x_0$, the hole density is uniform given by
\eq{peq} with $f_0=1/2$; the electrons are confined to
the grain boundary core by the grain boundary built-in electric field, leading
to a one-dimensional diffusion along it.  Solving the one-dimensional diffusion
equation leads to the electron density
\begin{align}
    n\GB(x) &= N_C e^{(-E_g+E\GB)/k_BT} e^{qV/k_BT} &\mathrm{for}~ x<x_0 \nonumber\\
            &= N_C e^{(-E_g+E\GB)/k_BT} e^{qV/k_BT} e^{-\frac{x-x_0}{L_n}} &\mathrm{for}~ x>x_0
    \label{ndecay}
\end{align}
where $L_n = \sqrt{2D_n L_{\mathcal{E}} / \mathcal{S}_n}$ ($D_n$: electron
diffusion coefficient) is the electron diffusion length, and
$L_{\mathcal{E}}=V_T/\mathcal{E}_y$ is the length scale of the confinement
($\mathcal{E}_y$: electric field transverse to the grain boundary in the bulk of
the $pn$ junction). $\mathcal{S}_n$ is the effective surface recombination
velocity in this case
\be
    \mathcal{S}_n = \sum_{k \in A,D} S_n^k \int_{0}^{E_g} \frac{\mathrm{d}E}{E_g}\ \frac{1}{1+ \frac{\bar
    p\GB(E)}{p\GB} + \frac{S^k_n \bar n\GB(E)}{S^k_p p\GB}}.
    \label{Sn}
\ee
The grain boundary dark current reads
\begin{align}
    J\GB(V) = \frac{\mathcal{S}_n}{d} N_C & e^{(-E_g+E\GB+qV)/k_BT} \nonumber\\
    &\times \left[x_0 + L_n\left(1 - e^{-\frac{L\GB-x_0}{L_n}} \right) \right],
    \label{jc2}
\end{align}
The integrals in
$\mathcal{S}_p$ have a similar interpretation as for the $n$-type grain boundary:
only the states with energies $E\GB \lesssim E \lesssim E_g - E\GB$ contribute
significantly to the recombination, as shown in Fig.~\ref{occupancy}(d).
The lower limit results from the fact that states below $E\GB$ are empty of
holes.  The upper limit is the energy at which electrons are emitted from the
defect state to the conduction band faster than holes relax from the valence
band to the defect state.  The integrands in $\mathcal{S}_n$ exhibit the same
shape as $\mathcal{S}_p$. In particular, the width of the integrand is still
given by \eq{delta}. Equation~(\ref{Sn}) therefore simplifies to
\be
    \mathcal{S}_p \approx S^A_n \Delta_A(\alpha) + S^D_n \Delta_D(\alpha).
\ee
Similarly to the $n$-type grain boundary, the occupancy of the gap states is
determined solely by holes and is therefore independent of the applied voltage,
shown in Fig.~\ref{occupancy}(c).

%================================================================
\subsection{Grain boundary recombination for $n\GB \propto p\GB$}
%================================================================
\label{pro}
As the applied voltage is increased above $E_g\Delta_k/q$, the minority carrier
density approaches the majority carrier density at the grain boundary. This
results in a rearrangement
of the gap states occupancies. However, we use the assumption that the grain
boundary charge does not change with voltage, as discussed in
Sec.~\ref{assumptions}, which leads to the constraint
\be
    1 = 1/E_g \int_0^{E_g} \mathrm{d}E\ f_D(E) + \alpha f_A(E).
        \label{conserve}
\ee
The change in occupancies keeps the area under the occupancy curves equal to its
equilibrium value, as shown by the dashed lines in
Figs.~\ref{occupancy}(a) and (c). More specifically,
occupancies above $E\GB$ increase while the ones below $E\GB$ decrease. These
changes lead to an increase of the number of states contributing to the
recombination as can be seen in Figs.~\ref{occupancy}(b) and (d).

There is no pinning of either quasi-Fermi level to $E_g/(1+\alpha)$ in this
regime.  We refer to Sec. III C of Ref.~\onlinecite{Gaury2016GB} for the
derivations in this case.  We can show that the constraint \eq{conserve} imposes
that the ratio of carrier densities remains constant along the grain boundary.
While this ratio was independent of voltage for the single acceptor+donor state
of Ref.~\onlinecite{Gaury2016GB}, this is not the case anymore. Assuming $p\GB =
\gamma(V) n\GB$, and solving a one-dimensional diffusion equation along the grain
boundary leads to the carrier densities
\begin{align}
\label{nc}
    n\GB(x) &= \frac{1}{\sqrt{\gamma}}n_i e^{qV/(2k_BT)} e^{-\frac{x}{L_n'}}\\
    p\GB(x) &= \sqrt{\gamma} n_i e^{qV/(2k_BT)} e^{-\frac{x}{L_n'}}.
\label{pc}
\end{align}
We find $\gamma(V)$ by solving \eq{conserve}. This
ratio gives the value of the plateau reached by the levels occupancy at energies
around midgap, as shown by the dashed lines in Figs.~\ref{occupancy}(a) and (c).
Because $n\GB$ and $p\GB$ dominate the level occupancy \eq{eq:fGBe} at these
energies, the value of the plateau is $1/(1+\gamma \beta_k)$, with $\beta_k =
S_p^k / S_n^k$. Considering (unrealistically) large values of applied voltage
such that the levels occupancy
is entirely independent of energy, we find that $\gamma$ converges to
\be
    \gamma = \frac{\alpha -1 + \sqrt{(1-\alpha)^2 + 4\alpha
    \beta_A/\beta_D}}{2\beta_A}.
\label{gamma}
\ee
In Eqs.~(\ref{nc}) and (\ref{pc}),
$L'_n=\sqrt{4D_nL_{\mathcal{E}}'/\mathcal{S}}$ is the diffusion length of electrons along the
grain boundary, and $L_{\mathcal{E}}'$ is the characteristic length
of the electric field transverse to the grain boundary.
The effective surface recombination velocity $\mathcal{S}$ reads
\be
\mathcal{S} = \sum_{k \in A,D} \frac{\gamma S_n^k S_p^k}{S_n^k+\gamma S_p^k} \int_{0}^{E_g}
\frac{\mathrm{d}E}{E_g}\ \frac{1}{1+ \frac{S_n^k \bar n\GB(E) + S_p^k \bar
p\GB(E)}{(S_n^k+\gamma S_p^k) \frac{n_i}{\sqrt{\gamma}} e^{qV/(2k_BT)}}}.
\label{S}
\ee
The integrand of \eq{S} now varies with voltage as shown by the dashed line in
Fig.~\ref{occupancy}(b). It can be shown that an approximation for the integral
is
\be
    \mathcal{S} \approx \sum_{k \in A,D} \frac{\gamma S_n^k S_p^k}{S_n^k+\gamma
    S_p^k} \left[\frac{qV}{E_g} - \frac{2k_BT}{E_g}\ln\left(\frac{\sqrt{\gamma}S_n^k}{S_n^k+\gamma
    S_p^k} \right) \right].
    \label{S2}
\ee
Gathering the above results leads to the grain boundary dark current
\be
    J\GB(V) = \frac{\mathcal{S}L'_n}{\sqrt{\gamma}d} n_i e^{V/(2V_T)} \left[1 -
    e^{-L\GB/L'_n}\right].
    \label{jc3}
\ee
This result is formally similar to the corresponding case for the single
acceptor+donor state
studied in Ref.~\onlinecite{Gaury2016GB}, yet the voltage dependence of the
effective surface recombination velocity is a key difference.
Taking $S_{n,p}^A=S_{n,p}^D$, \eq{jc3} is proportional to
$\sqrt{\alpha}/(1+\alpha)$.  This shows that
when the distribution of gap states is skewed towards either donors or
acceptors (e.g. big or small $\alpha$), the recombination current is diminished.
This reduction can be understood with the levels occupancy in
Fig.~\ref{occupancy}(c) (dashed line).  The recombination plotted in
Fig.~\ref{occupancy}(d) shows that the states around midgap contribute the most
to the recombination. These states correspond to the plateau of the levels
occupancy, which is approximately equal to $1/(1+\alpha)$ here. For large or
small $\alpha$, the plateau is far from $1/2$, which reduces the
probability that a hole \textit{and} an electron be captured together by a gap
state. This reduced probability leads to a reduction of the recombination
current.

In all three gap state configurations studied (single and continuum of
acceptor+donor, single donor), the high recombination regime exhibits a thermal
activation energy $E_g/2$ and an ideality factor of 2 (both typical of
recombination determined by electrons and holes equally). These characteristics
were observed in previous experimental work on Si $pn^+$ junctions aiming to
isolate the grain boundary recombination current~\cite{neugroschel1982effects}.

%============================================================
\subsection{Numerical verification of the analytical results}
%============================================================
The numerical tests to verify the accuracy of the results of this section are
presented in Fig.~\ref{4plotsc}. These results were obtained with
$S_{n,p}^A=S_{n,p}^{D}$.
For applied voltages below $E_g\Delta_k/q$, the current is given by the smallest value
between the $n$-type and $p$-type regimes. For higher values of applied voltage we
used the high recombination regime.
For all plots except Fig.~\ref{4plotsc}(b) we used $\rho_A/\rho_D=2$, which
leads to $E\GB=0.25~\rm eV$ ($p$-type grain boundary).
\begin{figure}
   \includegraphics[width=0.49\textwidth]{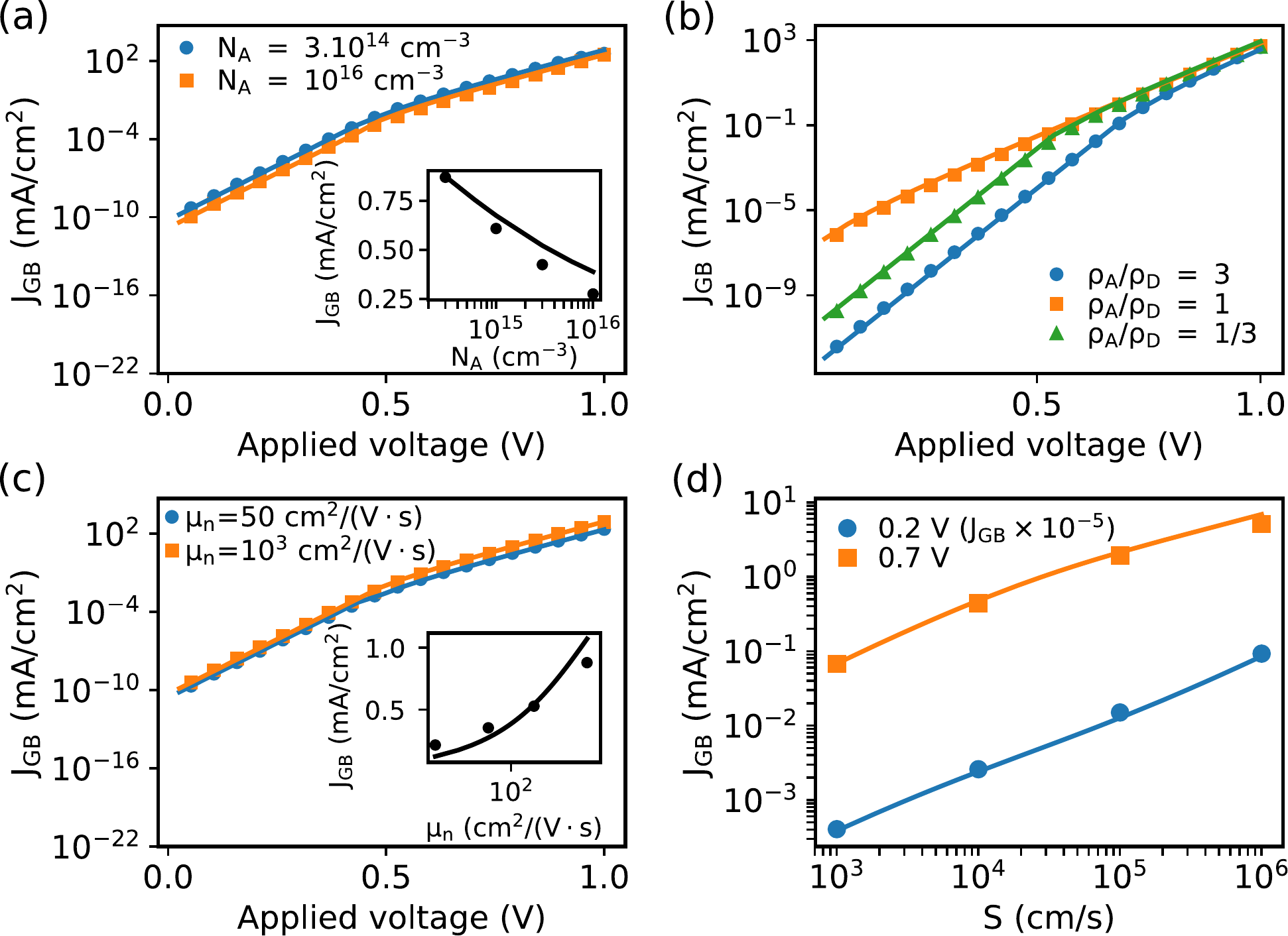}
    \caption{\label{4plotsc} Grain boundary recombination current characteristics
    $J\GB(V)$ for a continuum of donor and acceptor states with $\rho_A/\rho_D =
    2$, $N_A=10^{15}~\rm{cm^{-3}}$, $S_{n,p}^A=S_{n,p}^D=10^5~\rm{cm/s}$ and
    $\mu_n=320~\rm{cm^2/(V\cdot s)}$ unless specified otherwise. Symbols are
    numerical calculations, full lines correspond to analytical results. (a)
    $J\GB(V)$ varied with doping density. Inset: grain boundary recombination
    current as a function of doping density at $V=0.7~\rm{V}$. (b)
    $J\GB(V)$ varied with the ratio of acceptor to donor density of states. (c)
    $J\GB(V)$ varied with electron mobility. Inset:  grain boundary
    recombination current as a function of electron mobility for $V=0.7~\rm{V}$.
    (d) Grain boundary
    recombination current as a function of surface recombination velocity
    ($S_{n,p}^A=S_{n,p}^D$), at $V=0.2~\rm{V}$ (dots) and $V=0.7~\rm{V}$ (squares).}
\end{figure}

\bigskip
We vary the doping density in Fig.~\ref{4plotsc}(a). Before the change of slope,
the dark current is given by \eq{jc2} which exhibits a doping dependence mostly
via the width of the $n$-region $x_0$. The reduction of the slope reveals the
crossover to the high recombination regime ($n\GB \propto p\GB$) where the dark current is
given by \eq{jc3}. In this regime, the inset shows the predicted scaling in $N_A^{-1/4}$.

Fig.~\ref{4plotsc}(b) shows the grain boundary dark current for various ratios
$\rho_A/\rho_D$. In descending order, these correspond to grain boundary neutral
levels $E\GB \approx 0.38~\rm eV$, $0.75~\rm eV$ and $1.1~\rm eV$. The crossover
to the high recombination regime occurs when the occupancy of the gap
states starts to change significantly, that is for $qV > E_g \Delta_k$ (see
Fig.~\ref{occupancy}(b) and (d)).

The dependence of the dark current with electron mobility
is shown in Fig.~\ref{4plotsc}(c). At low voltage, this dependence is barely
visible but present. The dependence is weak in the $p$-type
regime because
both limiting cases of \eq{jc2}, $L_n\ll L\GB$ and $L_n\gg L\GB$, are
independent of mobility. The chosen set of parameters lies in between these
limits.  At high voltages we verify the square root scaling as shown by \eq{jc3}
in the limit $L_n' \ll L\GB$. 
Increasing electron mobility reduces the suppression of carrier den- sities away
from the maximum of recombination by increasing the electron diffusion length
$L^'_n$ . The increase of $L^'_n$, in turn, increases the
recombination along the grain boundary.

Fig.~\ref{4plotsc}(d) shows the scalings of the grain boundary dark current with
surface recombination velocity $S_{n,p}^{A,D}$. Because we used equal surface
recombination velocities for the donor and acceptor states, and electrons and
holes, the effective recombination velocities and $S_{n,p}^{A,D}$ are
proportional.
The notable feature of this plot is the $\sqrt{\mathcal{S}}$ scaling of the
grain boundary recombination current obtained at high voltage. This feature only
appears at high recombination velocities as it requires $L_n' \ll L\GB$, as
shown by \eq{jc3}. In the opposite limit one recovers a linear scaling in $\mathcal{S}$.

\section{Open-circuit voltage}
\label{secVoc}
We now consider a charged grain boundary under illumination and derive relations
for the open-circuit voltage $V_{\rm oc}$.  We assume that around $V_{\rm oc}$
the current-voltage relation under illumination is given by the sum of the short
circuit current  $J_{\rm sc}$ and the dark current (see Sec.~V of
Ref.~\onlinecite{Gaury2016GB} for a discussion on the validity of this
assumption). The dark current is the sum of the grain boundary dark current
(derived in Sec.~III and Sec.~IV) and the bulk recombination current.  We use
the results of Sec.~IV of Ref.~\onlinecite{Gaury2016GB} for the bulk
recombination whenever necessary.

\bigskip
Neglecting the bulk recombination and the non-exponential voltage dependences in
the grain boundary dark currents, we can write down explicit forms for the
open-circuit voltage associated with the grain boundary recombination. The
general form of the open-circuit voltage reads
\be
\label{voc}
    qV_{\rm oc}^{\rm GB} = nE_a + nk_BT \ln\left( \frac{d J_{\rm sc}}{S\lambda N}
    \right),
\ee
where $S$ is a surface recombination velocity, $\lambda$ is a length
characteristic of the physical regime, $N$ is an effective density of states, $E_a$ is
an activation energy, $V$ is the applied voltage and $n$ is an ideality factor.
Even though \eq{voc} is not mathematically correct in all cases (because it
neglects non-exponential voltage dependence), it captures the
dominant scalings for the physical parameters and should give the reader an
intuition for how these parameters impact $V_{\rm oc}$. The parameters entering
\eq{voc} are shown in Table~\ref{form} for all grain boundary configurations.

\subsection{Single donor state}
We begin with the single donor state in the gap of the absorber
material. The parameters in \eq{voc} for this case are given in the first row of
Table~\ref{form}. Fig.~\ref{Voc_d} shows comparisons between the numerically computed
$V_{\rm oc}$, and the values obtained with the numerically computed
$J_{\rm sc}$ and the analytic forms of the dark current.

\bigskip
These results differ somewhat from the single donor+acceptor defect case of
Ref.~\onlinecite{Gaury2016GB}. As we discussed in Sec.~\ref{3D}, the single donor
state with common material parameters for CdTe exhibits increasing grain
boundary dark currents with doping at high voltages.
As a result, $V_{\rm oc}$ decreases with doping as shown in
Fig.~\ref{Voc_d}(a).
Figure~\ref{Voc_d}(b) shows that the open-circuit voltage as a function of defect
energy level is not symmetrical from midgap. In fact, for $E\GB \gtrsim
1.2~\rm eV$ the open-circuit voltage is given by the grain boundary
recombination current of the $n$-type regime \eq{j1}, while for lower $E\GB$ values,
the open-circuit voltage is given by the regime $S_n n\GB \approx (1-f_0)
S_pp\GB$ and \eq{lim1} (high recombination regime). The plot shows that donor
states close to the band edge (i.e. $n$-type grain boundaries) are more
favorable to $V_{\rm oc}$. These states are less easily accessible to holes than
over states, hence reducing the probability for recombination.
Figure~\ref{4plots_D}(b) shows that, for a given applied voltage, the amplitude
of the grain boundary recombination current of the $n$-type regime is smaller
than the one of the high recombination regime (hence the larger $V_{\rm oc}$ in
the first case).  Figure~\ref{Voc_d}(c) shows the dependence with electron mobility for
values higher than $10~\rm cm^2/(V\cdot s)$.  Under the chosen conditions the
grain boundary recombination current depends weakly on mobility.  Finally the
logarithmic dependence on surface recombination velocity of \eq{voc} is shown in
Fig.~\ref{Voc_d}(d).
\begin{figure}
    \includegraphics[width=0.49\textwidth]{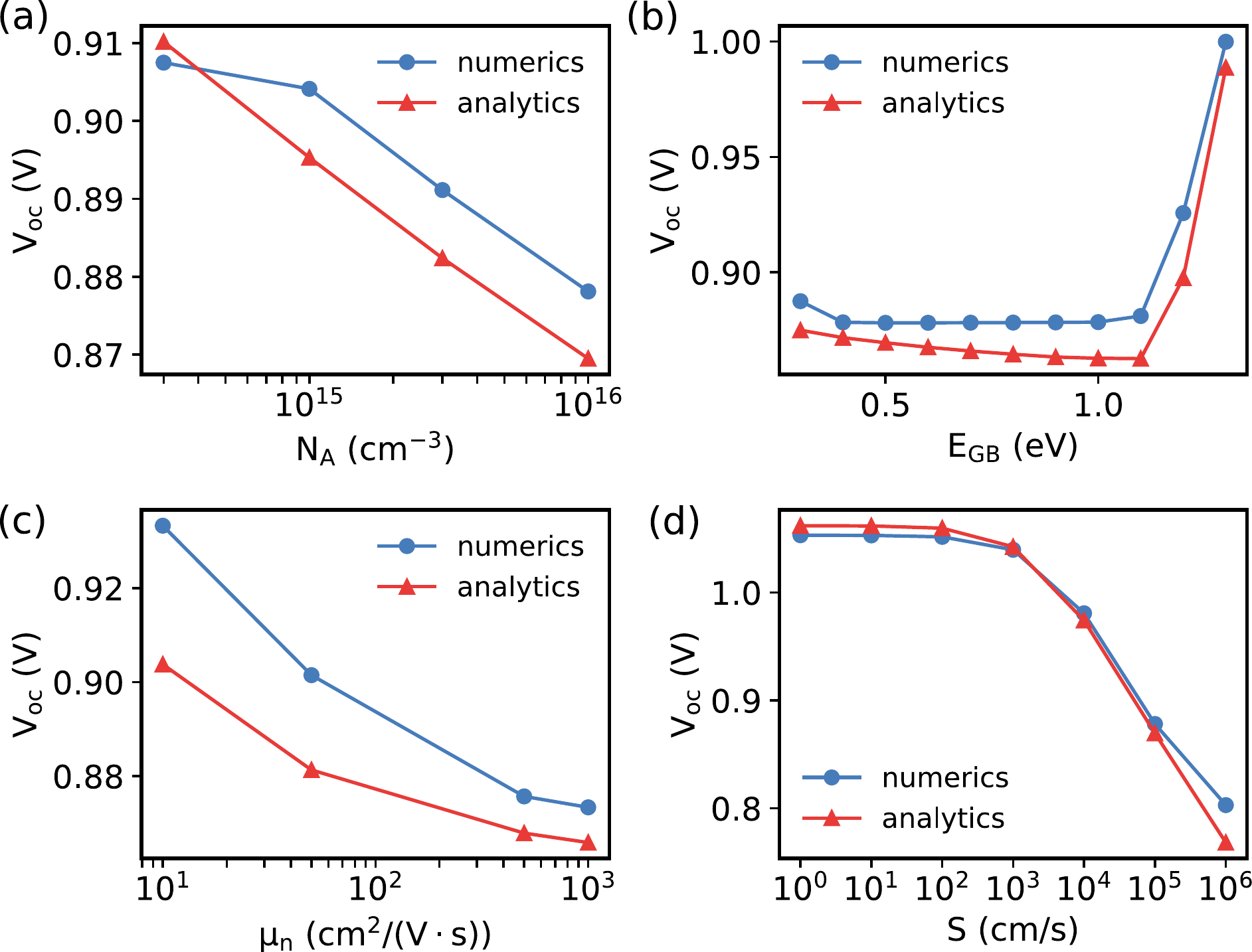}
    \caption{\label{Voc_d} Open-circuit voltage for the system described in
    Fig.~\ref{system}(a) with a single donor state at $E\GB=0.5~\rm eV$, with
    $\tau_n=100~\rm ns$,
     under a photon flux $10^{21}~\rm m^{-2}\cdot s^{-1}$.
    The absorption length is $2.3\times10^4~\rm cm^{-1}$.
    The electron mobility is $320~\rm cm^2/(V\cdot s)$,  $N_A=10^{16}~\rm
    cm^{-3}$ and $S_n=S_p=10^5~\rm cm/s$ unless specified
    otherwise. Numerical data are in blue (dots) and
    analytical predictions are in red (triangles). (a) $V_{\rm oc}$ as a
    function of doping density. (b) $V_{\rm oc}$ as a function of the defect
    state energy level.
    (c) $V_{\rm oc}$ as a function of electron mobility. (d) $V_{\rm oc}$ as a
    function of surface recombination velocity ($S_n=S_p$).}
\end{figure}

\bigskip
A key difference between this single donor case and the donor+acceptor case of
Ref.~\onlinecite{Gaury2016GB} is the amplitude of the grain boundary
recombination in the high recombination regime. In the single donor state of the
present work, the electron surface recombination velocity is effectively reduced by the
factor $1-f_0$ (which can be on the order of $10^{-3}$ in the regime of large
defect density of states) as can be seen in the expressions of the carrier
densities Eqs.~(\ref{n'}) and (\ref{px}), as well as in the recombination itself
\eq{Rd}.  As a result, for intermediate values of bulk lifetime ($\approx 10~\rm
ns$), the bulk recombination current is of the same order of magnitude as the
\begin{figure}
    \includegraphics[width=0.49\textwidth]{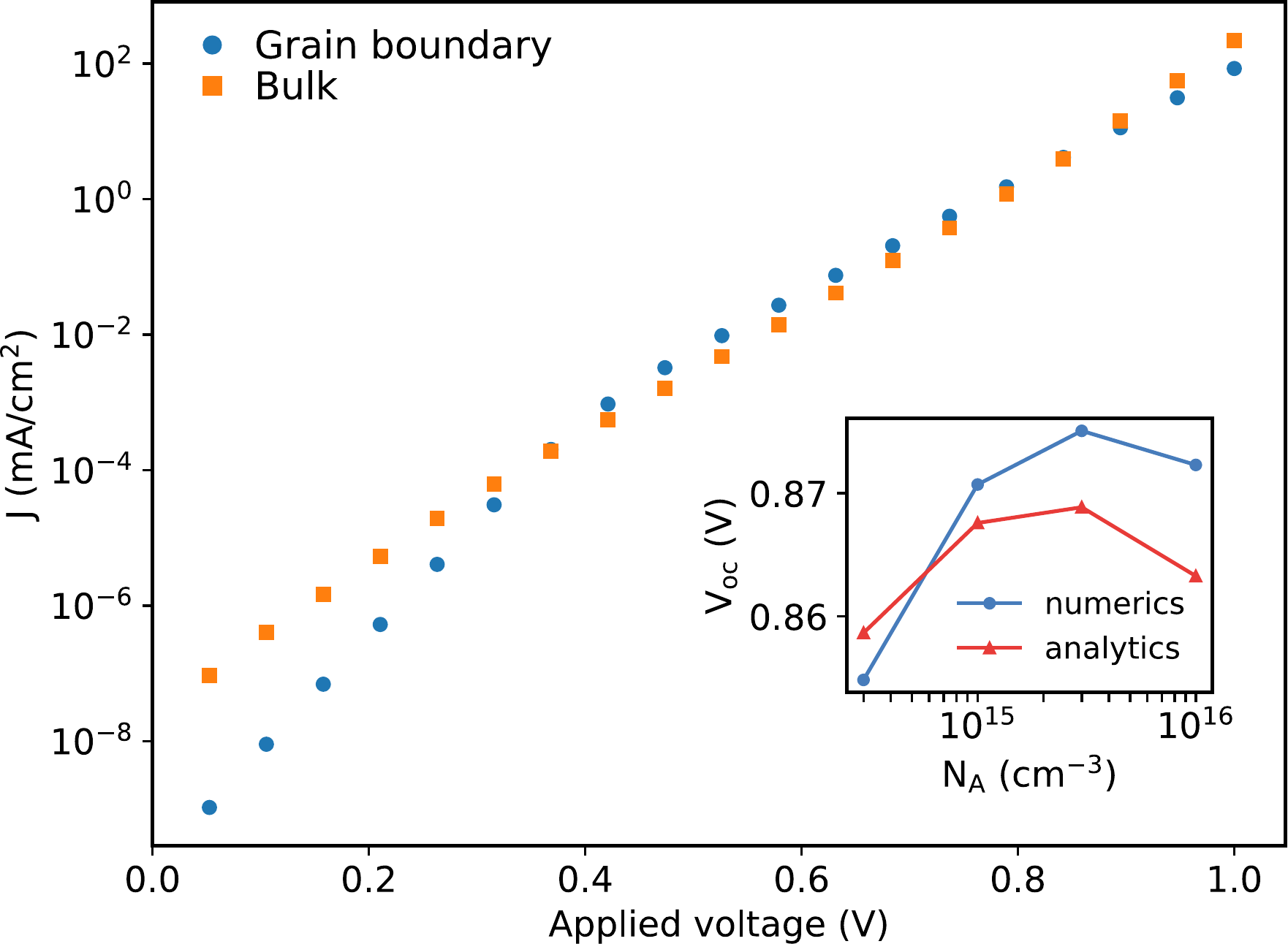}
    \caption{\label{bulk} Recombination current as a function of voltage for the
    grain boundary (blue dots) and the bulk (orange squares) obtained with a
    bulk lifetime $\tau_n=10~\rm ns$. Other parameters: $N_A=10^{15}~\rm
    cm^{-3}$, $S_{n,p}=10^5~\rm cm/s$, $\mu_n=320~\rm cm^2/(V\cdot s)$,
    $E\GB=0.5~\rm eV$.  Inset: open-circuit voltage for our system  under a
    photon flux $10^{21}~\rm m^{-2}\cdot s^{-1}$.  The absorption length is
    $2.3\times10^4~\rm cm^{-1}$.  }
\end{figure}
grain boundary recombination current with $S_{n,p}=10^5~\rm cm/s$, as shown in
Fig.~\ref{bulk}. In this example, reducing the doping density
increases the bulk recombination (because the width of the $pn$ junction
depletion region increases) which now dominates over the grain boundary
contribution. The inset of Fig.~\ref{bulk} shows that the resulting open-circuit
voltage increases with doping, contrary to the behavior shown in
Fig.~\ref{Voc_d}(a) for which $\tau=100~\rm ns$.

\subsection{Continuum of gap states}
We now turn to the continuum of acceptor and donor states. The parameters in
\eq{voc} are given in the last row of Table~\ref{form}. Note that in the high
recombination regime ($n\GB = \gamma p\GB$), \eq{voc} applies when neglecting the
linear voltage dependence of the effective surface recombination velocity
$\mathcal{S}$ (see \eq{S2}). Despite this approximation, \eq{voc} provides the
correct overall scalings with doping density, distribution of states, mobility
and surface recombination velocities. Fig.~\ref{Voc_c} shows the comparisons of the
simulated open-circuit voltages with our analytical results.

\begin{figure}
    \includegraphics[width=0.49\textwidth]{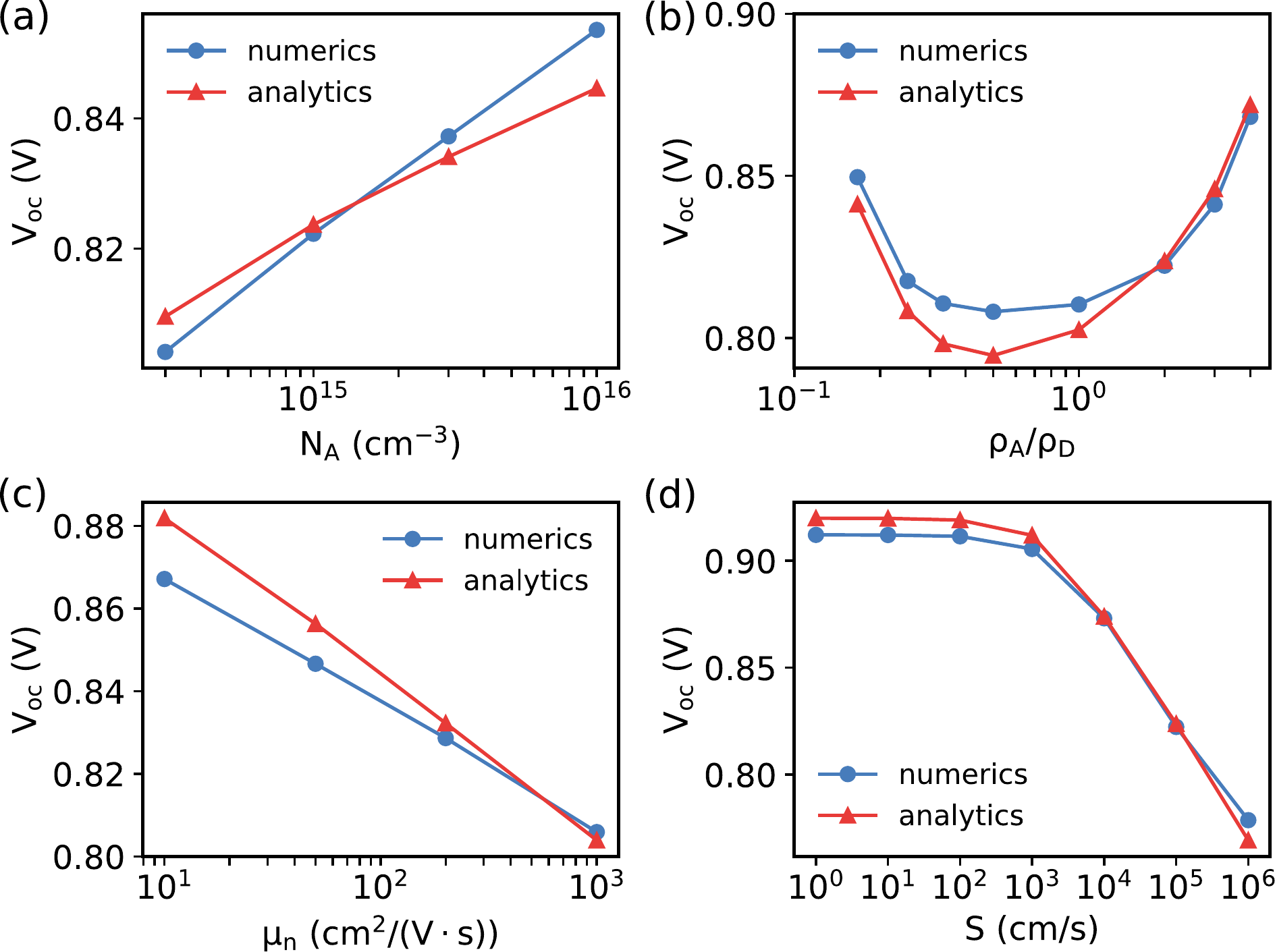}
    \caption{\label{Voc_c} Open-circuit voltage for the system described in
    Fig.~\ref{system}(a) with a continuum of donor and acceptor states with $\rho_A/\rho_D =
    2$, under a photon flux $10^{21}~\rm m^{-2}\cdot s^{-1}$.
    The absorption length is $2.3\times10^4~\rm cm^{-1}$.
    The electron mobility is $320~\rm cm^2/(V\cdot s)$,  $N_A=10^{15}~\rm
    cm^{-3}$ and $S_n=S_p=10^5~\rm cm/s$ unless specified
    otherwise. Numerical data are in blue (dots) and
    analytical predictions are in red (triangles). (a) $V_{\rm oc}$ as a
    function of doping density. (b) $V_{\rm oc}$ as a function of the ratio of
    acceptor and donor densities of states.
    (c) $V_{\rm oc}$ as a function of electron mobility. (d) $V_{\rm oc}$ as a
    function of surface recombination velocity ($S_n=S_p$).}
\end{figure}

\bigskip
The continuum of states studied here has lots of features similar to the single
acceptor+donor state studied in Ref.~\onlinecite{Gaury2016GB}. In particular we find the same
scalings of $V_{\rm oc}$ with doping, mobility and surface recombination
velocity shown in Fig.~\ref{Voc_c}(a),
(c) and (d).
A difference with Ref.~\onlinecite{Gaury2016GB} is the U-shape dependence
 of $V_{\rm oc}$ with the ratio $\rho_A/\rho_D$, presented in
Fig.~\ref{Voc_c}(b).  The open-circuit voltage now varies with $\rho_A/\rho_D$
(i.e. with the effective single state value) even for intermediate values of the
ratio. This is because the grain boundary recombination current in the high
recombination regime depends on this ratio via $\gamma$ as shown by \eq{gamma}.

Figure~\ref{Voc_c}(b) shows that gap state configurations with ratios
$\rho_A/\rho_D$ away from 1 give better $V_{\rm oc}$ values.  Note that we have
assumed equal values of short circuit current $J_{\rm sc}$ for all values of
grain boundary parameters.  This will certainly not be the case in practice.
Indeed, for grain boundaries which do not undergo type inversion (e.g. $p$-type
grain boundaries), we find that the short circuit current density is decreased.
Therefore, only gap state configurations with more donor states will be
beneficial for photovoltaic performance for the model configurations
studied in this paper.

%==================================
\section{Conclusion}
%==================================

We generalized the physical descriptions associated with the microscopic charge
transport and recombination of Ref.~\onlinecite{Gaury2016GB} to two additional
configurations of gap states: a single donor state and a continuum of donor and
acceptor states. In this work we derived the corresponding analytic expressions
for the grain boundary dark recombination current. We found that all these
configurations share three similar regimes describing the grain boundary dark
recombination current as a function of voltage ($n$-type, $p$-type and
high-recombination). However, they exhibit differences in the amplitude of the
subsequent recombinations. Mixtures of acceptor and donor states (single level
and continuum) result in enhanced recombination that reduces the open-circuit
voltage for a wide range of bulk lifetimes. The amplitudes of recombination for
the single donor state are lower, and commensurate with the bulk recombination
for intermediate bulk lifetimes. From this work and Ref.~\onlinecite{Gaury2016GB}, we
observed that a larger concentration of donor states compared to acceptor
states improve the device open-circuit voltage for a fixed value of the short
circuit current density.

Nanoscale imaging and spectroscopy combined with first principles calculations
can now determine, at least in principle, the electronic configuration of grain
boundaries~\cite{Yanfa2015}. In turn, this knowledge can be used within the
framework developed here to obtain quantitative predictions of device
open-circuit voltage. In this way, our work provides a bridge between nanoscale
characterization and macroscopic device response. Finally, our approach and the
physical descriptions of grain boundaries presented here extend beyond CdTe or
Cu(In,Ga)Se$_2$ technologies. For example, our approach could be applied to
grain boundaries with upward band banding resulting from negatively charged
boundaries in $p$-type materials or alternatively, positively charged boundaries
in $n$-type materials.

\begin{acknowledgements}
B.~G. acknowledges support under the Cooperative Research
Agreement between the University of Maryland and the National Institute of
Standards and Technology Center for Nanoscale Science and Technology, Award
70NANB14H209, through the University of Maryland.
\end{acknowledgements}

\begin{appendices}
%=============================================================================
\section{Conditions for large defect density regime}
%=============================================================================
\label{pinning}
We derive the minimal defect densities for which $Q\GB/(qN\GB) \ll 1$ for the single
donor defect state and the continuum of defect states. Because the defect
statistics in thermal equilibrium are different in each case (see \eq{QGB1} and
\eq{QGB2} respectively), we must derive two different criteria.

\bigskip
We start with the single donor state. The condition $Q\GB/(qN\GB) \ll 1$ requires
that $f\GB \approx 1$. We specify this requirement by imposing that the Fermi
level lies at least $3k_BT$ above the defect level at the grain boundary,
\be
    E_F + qV_{\rm GB}^0 - E\GB > 3k_BT.
\ee
This condition together with \eq{VGB} imposes
\be
    1-f_0 > \frac{1}{1+e^3},
\ee
where $f_0=(1+\exp[(E\GB-E_F)/k_BT])^{-1}$.
Using a depletion approximation and $qV_{\rm GB}^0 = E\GB-E_F + 3k_BT$, the
charge in the depleted regions surrounding the grain boundary is
\be
    Q = \sqrt{8\epsilon qN_A V_{\rm GB}^0} \approx \sqrt{8\epsilon N_A(E\GB-E_F
    + 3k_BT)}.
    \label{Q}
\ee
Equating \eq{QGB1} and \eq{Q} leads to the critical defect density of states
\be
    N_{\rm GB}^{\rm crit} = \frac{1+e^3}{q} \sqrt{8\epsilon N_A(E\GB-E_F + 3k_BT)}.
\ee

For the case of the continuum of acceptor and donor states, the large defect
density of states corresponds to the pinning of the Fermi level to
the neutral point of the gap states distribution. We will consider the large
defect density regime when the distance between $E_F$ and $E\GB$ is smaller than
$k_BT$. Using a depletion approximation and $qV_{\rm GB}^0 = E\GB-E_F-k_BT$, the
charge in the depleted regions around the grain boundary is
\be
    Q \approx \sqrt{8\epsilon N_A (E\GB-E_F-k_BT)}.
    \label{Q2}
\ee
Equating Eqs.~(\ref{QGB}) and (\ref{Q2}) leads to the critical value
\be
    \rho_D^{\rm crit} = \frac{1}{q} \frac{\sqrt{8\epsilon N_A
    (E\GB-E_F-k_BT)}}{
    k_BT \ln\left(\frac{1+e^{(E_g-E\GB+k_BT)/k_BT}}{1+e^{(-E\GB+k_BT)/k_BT}} \right)
    -\alpha E_g }
    \label{critt}
\ee
where $E\GB = E_g/(1+\alpha)$ with $\alpha = \rho_A/\rho_D$. This critical donor
density of states depends on the ratio $\alpha$ considered. Also note that
because the denominator of \eq{critt} depends on energy,
$\rho_D^{\rm crit}$ is a density of states per energy unit
(expressed in $\rm m^{-2}\cdot eV^{-1}$).

% we use the mapping
% onto a single state problem where the charge at the grain boundary is given by
% \eq{QGB2}. Here we simply use the criterion derived in Appendix~A of
% Ref.~\onlinecite{Gaury2016GB} and find the critical donor density
% \be
%     \rho_D^{\rm crit} = \frac{1}{q} \left(\frac{e+1}{e-1} \right) \frac{\sqrt{8qN_A(E\GB -
%     E_F)}}{E_g},
% \ee
% where $E\GB = E_g/(1+\alpha)$ with $\alpha = \rho_D/\rho_A$. This critical donor
% density of states depends on the ratio $\alpha$ considered. Also note that because of the
% division by $E_g$, $\rho_D^{\rm crit}$ is a density of states per energy unit
% (expressed in $\rm m^{-2}\cdot eV^{-1}$).

%==========================================================
\section{Condition for nearly flat hole quasi-Fermi level}
%==========================================================
\label{flat}
We specify the domain of validity of the assumption of flat hole quasi-Fermi
level. In the main text we consider $E_{F_p}=E_F$ when variations of
$E_{F_p}$ across the grain boundary are smaller than $k_BT$.  An expansion of
$E_{F_p}$ across the grain boundary yields
\be
    E_{F_p} = E_F - \left|\frac{\partial E_{F_p}}{\partial y}\right| \delta y,
\ee
where the gradient of $E_{F_p}$ at the grain boundary depends on the regime
considered (e.g. $n$-type grain boundary). The single donor state and the
continuum of defect states require formally the same condition. The relevant
surface recombination velocity must be used in each case. We are able to derive
such a criterion only for an $n$-type grain boundary but found that it applies
well also in other regimes. In what follows we focus on the single donor state.

In the regime $S_pp\GB \ll S_n\bar n\GB$, the gradient of $E_{F_p}$ is obtained by
integrating the continuity equation for holes across the grain boundary over an
infinitely small distance,
\be
    \left|\frac{\partial E_{F_p}}{\partial y}\right| = q\frac{S_p}{2\mu_p}.
\ee
Assuming that the variation of $E_{F_p}$ across the grain boundary follows that of the
electrostatic potential, the distance across the grain boundary where
$E_F-E_{F_p} < k_BT$ is given by a depletion approximation $\delta y =
\sqrt{2\epsilon V_T/(qN_A)}$. The assumption of flat $E_{F_p}$ is therefore valid
for
\be
    \frac{S_p}{2\mu_p}\sqrt{\frac{2\epsilon}{qV_TN_A}} < 1.
    \label{criterion1}
\ee
For the continuum of states one would use $\mathcal{S}_p$ instead.

%=================================================================
\section{Derivations for $p$-type donor defect state}
%=================================================================
\label{derivations:nsmall}
Using the energy scale and definitions of
Fig.~\ref{donor_ptype}(d), the carrier densities at the grain boundary are given by
\begin{align}
    \label{nn}
    n\GB(x) &= N_C e^{(E_{F_n}(x) + q\phi\GB(x) -E_g)/k_BT},\\
    p\GB(x) &= N_V e^{(-E_{F_p}(x) - q\phi\GB(x))/k_BT},
    \label{pp}
\end{align}
where $\phi\GB$ is the electrostatic potential at the grain boundary.  The zero
of electrostatic potential is at the $p$-contact away from the grain
boundary. In the $p$-type donor state, the electrostatic potential
is uniform along the grain boundary in the bulk of the $pn$ junction and is
independent of voltage. The value of $\phi\GB$ is determined by the grain
boundary built-in potential at thermal equilibrium,
\be
    \label{e}
    q\phi\GB = qV_{\rm GB}^0,
\ee
where $V_{\rm GB}^0$ is given by \eq{vgbD}.
The grain boundary recombination is maximum at $x_0$, where $x_0$ is the point
where the carrier densities in the grain interior are equal. We consider that
the grain interior recombination is peaked at the same point.
Using a depletion approximation in the depletion region of the $pn$ junction in
the grain interior, we find that $n=p=n_i$ at
\be
    x_0 = \sqrt{\frac{2\epsilon V_{\rm bi}}{qN_A}}
    \left[1 - \sqrt{1 - \frac{V_T}{V_{\rm bi}} \ln\left(\frac{N_D}{n_i} \right)} \right],
    \label{x0}
\ee
where $V_{\rm bi}$ is the $pn$ junction built-in potential (the dependence of
$x_0$ on applied voltage is weak and can be neglected). Using the potential
Eq.~(\ref{e}) in Eqs.~(\ref{nn}) and (\ref{pp}), and assuming that
$E_{F_p}=E_F$~(see justification above \eq{assumption_efp_flat}), we obtain
expressions for the carrier densities beyond $x_0$,
\begin{align}
    \label{d}
    n\GB(x \geq x_0) &= \frac{N_C}{1-f_0} e^{(-E_g+E\GB)/k_BT}e^{(E_{F_n}(x)-E_F)/k_BT},\\
    p\GB(x \geq x_0) &= (1-f_0)N_V e^{-E\GB/k_BT}.
    \label{dd}
\end{align}
Despite the above formulation, these relations are valid only for $x\gg x_0$.
We extend their domain of validity to $x=x_0$ for the purpose of calculating the
recombination, where the errors we make here on $n\GB$ and $p\GB$ cancel out.

For $x \geq x_0$ we use the continuity equation for electrons to obtain
$E_{F_n}$,
\be
    \frac{\partial J_{n,x}}{\partial x} + \frac{\partial J_{n,y}}{\partial y} =
    S_n (1-f_0) n\GB  \delta(y-y\GB) + R_{\rm bulk}(y),
    \label{continuity1}
\ee
where $R_{\rm bulk}$ is the bulk recombination and the electron current
component along the grain boundary is given by
\be
    J_{n,x}(x,y) =  \mu_n n\GB(x) e^{-y/L_{\mathcal{E}}} \frac{\partial E_{F_n}}{\partial x}(x).
    \label{jx}
\ee
In the above equation we assumed that the electron density across the grain boundary decays
as $e^{-y/L_{\mathcal{E}}}$, where
\be
    L_{\mathcal{E}} = V_T\sqrt{2 \epsilon /(qN_AV_{\rm GB}^0)}
    \label{ey}
\ee
is the characteristic length associated with the electric field transverse to the
grain boundary in the bulk region. This exponential decay assumes that $E_{F_n}$
is flat around the grain boundary, which coincides with the fact that the
currents going to the grain boundary are small. Integrating
Eq.~(\ref{continuity1}) in the $y$-direction around the grain boundary leads to
\be
2L_{\mathcal{E}} \mu_n   k_BT\frac{\partial^2}{\partial x^2}
    \left[e^{E_{F_n}/k_BT}\right] =
    q S_n (1-f_0) e^{E_{F_n}/k_BT},
    \label{ugly}
\ee
where we neglected the currents in the $y$-direction at the end of the grain
boundary depletion region, and the bulk recombination.  We introduce the
effective diffusion length $L_n = \sqrt{2D_nL_{\mathcal{E}}/(S_n (1-f_0))}$,
where $D_n= k_BT\mu_n/q$ is the electron diffusion constant, to rewrite Eq.~(\ref{ugly}) as
\be
\frac{\partial^2}{\partial x^2} \left[e^{E_{F_n}/k_BT}\right] =
     \frac{1}{L_n^2} e^{E_{F_n}/k_BT}.
     \label{eqefn}
\ee
Considering that $E_{F_n}=E_F+qV$ at $x=x_0$, and neglecting the diverging part of
the solution of Eq.~(\ref{eqefn}), we obtain
\be
    E_{F_n}(x \geq x_0) = E_F+qV - k_BT \frac{x-x_0}{L_n}.
    \label{efnsol}
\ee
We verify the accuracy of Eq.~(\ref{efnsol}) in Fig.~\ref{2plots2}(b) (black
dashed curve).

Inserting Eq.~(\ref{efnsol}) into Eq.~(\ref{d}) yields
the electron density given in the main text
\be
    n\GB(x \geq x_0) = \frac{N_C}{1-f_0} e^{(-E_g+E\GB+qV)/k_BT}
    e^{-\frac{x-x_0}{L_n}}.
    \label{ndecayhere}
\ee
Because $S_nn\GB \ll S_p\bar p\GB$, the recombination at the grain boundary reads
\be
    R\GB(x\geq x_0) = S_n (1-f_0)n\GB(x\geq x_0).
\ee
We integrate over the length of the grain boundary to obtain the recombination
current
\be
    J\GB(V) = \frac{S_n N_C}{d} e^{(-E_g+E\GB + qV)/k_BT} L_n \left[1 -
    e^{-\frac{L\GB-x_0}{L_n}}\right].
    \label{JGBnsmall}
\ee
Equation~(\ref{JGBnsmall}) is the general result in the case $S_nn\GB \ll
S_p\bar p\GB$.
\begin{figure}
   \includegraphics[width=0.48\textwidth]{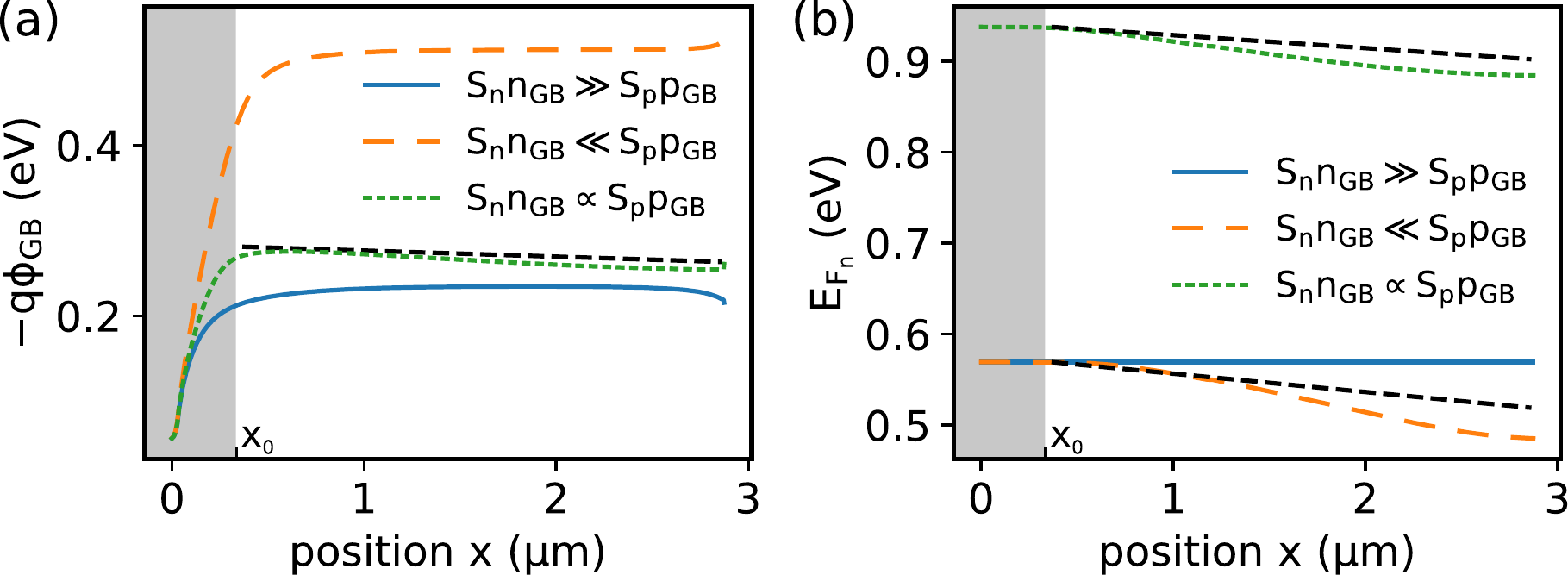}
    \caption{\label{2plots2} Numerical data computed along the grain boundary
    for the parameters of Fig.~\ref{2plots} with mobilities
    $\mu_n=10~\rm cm^2/(V\cdot s)$, $\mu_p=40~\rm cm^2/(V\cdot s)$. (a) Electrostatic
    potential. The dark dashed line corresponds to Eq.~(\ref{eq:phiGB}). (b) Electron
    quasi-Fermi level. The dark dashed lines correspond to Eq.~(\ref{eq:efn}) (upper)
    and Eq.~(\ref{efnsol}) (lower).
    }
\end{figure}

%=================================================================
\section{Derivations for donor state in the high recombination regime}
%=================================================================
\label{efndemo}
Here we provide the derivations of the analytical results presented in
Sec.~\ref{n_prop_p}. In the high recombination regime there is a constant $k$ such
that $S_p p\GB = k S_n n\GB$ along the grain boundary. In addition $n\GB
p\GB=n_i^2e^{(E_{F_n}-E_{F_p})/k_BT}$, so that we get
\be
    n\GB(x) = \sqrt{\frac{S_p}{kS_n}}n_i e^{(E_{F_n}(x)-E_{F_p}(x))/(2k_BT)}.
    \label{start}
\ee
Similarly to the $p$-type case, the recombination is peaked at $x_0$ and decays
after that point so we focus the derivation beyond $x_0$.

From here on the derivation of $E_{F_n}$ follows the exact same steps as
Appendix~\ref{derivations:nsmall} starting with the continuity equation:
\be
    \frac{\partial J_{n,x}}{\partial x} + \frac{\partial J_{n,y}}{\partial y} =
    \frac{k}{1+k} S_n n\GB \delta(y-y\GB) + R_{\rm bulk}(y)
    \label{continuity}
\ee
where $L'_{\mathcal{E}} = \sqrt{2\epsilon V_T/(qN_A)}$. $L'_{\mathcal{E}}$
is the characteristic length associated with the electric field transverse to
the grain boundary. $R_{\rm bulk}$ is again the bulk recombination. Because the
grain boundary built-in potential is not uniform in this regime, the transverse
electric field depends on the location along the
grain boundary. While $L'_{\mathcal{E}}$ does not correspond to a precise field,
we find that it accurately determines the slopes of the electron quasi-Fermi
level and the electrostatic potential along the grain boundary. The electron
current is still given by Eq.~(\ref{jx}) with the change of $L_{\mathcal{E}}$
for $L'_{\mathcal{E}}$. Integrating Eq.~(\ref{continuity}) around the grain
boundary leads to
\be
4L'_{\mathcal{E}} \mu_n   k_BT\frac{\partial^2}{\partial x^2}
    \left[e^{\frac{E_{F_n}-E_{F_p}}{2k_BT}}\right] =
    q \frac{k}{1+k}S_n e^{\frac{E_{F_n}-E_{F_p}}{2k_BT}},
    \label{i}
\ee
where we neglected the currents in the $y$-direction at
the end of the grain boundary depletion region, and the bulk recombination.
We introduce the effective diffusion length
$L'_n = \sqrt{4D_n L'_{\mathcal{E}}(1+k)/(kS_n)}$, and assume that
$E_{F_p}=E_F$~[30] to rewrite Eq.~(\ref{i}) as
\be
\frac{\partial^2}{\partial x^2} \left[e^{E_{F_n}/(2k_BT)}\right] =
     \frac{1}{{L'_n}^2} e^{E_{F_n}/(2k_BT)}.
\ee
Considering that $E_{F_n}=E_F+qV$ at $x=x_0$ we obtain
\be
    E_{F_n}(x\geq x_0) = E_F+qV - 2k_BT \frac{x-x_0}{L'_n}.
    \label{eq:efn}
\ee
Since $k S_nn\GB = S_pp\GB$, we can equate Eqs.~(\ref{nn}) and (\ref{pp}) to get
\be
    E_{F_n}(x) = -2q\phi\GB(x) - E_F -E_g-k_BT\ln\left(k\frac{S_nN_C}{S_pN_V}
    \right),
\ee
which yields the electrostatic potential along the grain boundary
\be
    \phi\GB(x) = k_BT\frac{x-x_0}{L'_n} -E_F -q\frac{V}{2} -
    k_BT\ln\left(\frac{n_i}{N_V}\sqrt{\frac{kS_n}{S_p}} \right).
    \label{eq:phiGB}
\ee
Comparisons of Eq.~(\ref{eq:efn}) and Eq.~(\ref{eq:phiGB}) with numerical data
are shown in Figs.~\ref{2plots2}(a) and \ref{2plots2}(b) respectively (dotted
green curves). We see that the numerically computed potential and electron
quasi-Fermi level are not linear over the entire length of the
grain boundary, however the analytical results give a good approximation of the
slopes near the depletion region.

Inserting Eqs.~(\ref{eq:efn}) and (\ref{eq:phiGB}) into the densities
Eqs.~(\ref{nn}) and (\ref{pp}) yields the densities given in Sec.~\ref{n_prop_p}.
These densities yield the grain boundary recombination
\be
    R\GB(x>x_0) = \frac{\sqrt{kS_nS_p}}{1+k} n_i e^{qV/(2k_BT)}
    e^{-\frac{x-x_0}{L_n'}}.
\ee
Integrating the recombination over the length of the grain
boundary gives the recombination current
\be
    J\GB(V) = \frac{\sqrt{kS_nS_p}L'_n}{(1+k)d} n_i e^{V/(2V_T)} \left[1 -
    e^{-L\GB/L'_n}\right].
    \label{jgbn=p}
\ee
Equation~(\ref{jgbn=p}) is the general result in the case $kS_nn\GB =
S_pp\GB$.

The constant $k$ can be specified considering that the occupancy of the grain boundary
defect level remains equal to its thermal equilibrium value $f_0$. We thus find
\be
    k \approx 1-f_0,
\ee
assuming that $f_0 \lesssim 1$ because of the high defect density of states.

\end{appendices}

\bibliography{ref}

\end{document}